# Infrared Imaging using thermally stable HgTe/CdS nanocrystals

Huichen Zhang[1$], Yoann Prado[1$,*], Rodolphe Alchaar[1], Henri Lehouelleur[2], Mariarosa Cavallo[1], Tung Huu Dang[1,3], Adrien Khalili[1], Erwan Bossavit[1,4], Corentin Dabard[2], Nicolas Ledos[1], Mathieu G Silly[4], Ali Madouri[5], Daniele Fournier[1], James K. Utterback[1], Debora Pierucci[1], Victor Parahyba[6], Pierre Potet[6], David Darson[3], Sandrine Ithurria[2], Bartłomiej Szafran[7], Benjamin T. Diroll[8], Juan I. Climente[9], Emmanuel Lhuillier[1*]

[1] Sorbonne Université, CNRS, Institut des NanoSciences de Paris, INSP, F-75005 Paris, France.
[2] Laboratoire de Physique et d'Etude des Matériaux, ESPCI-Paris, PSL Research University, Sorbonne Université, CNRS UMR 8213, 10 rue Vauquelin 75005 Paris, France.
[3] Laboratoire de Physique de l'Ecole normale supérieure, ENS, Université PSL, CNRS, Sorbonne Université, Sorbonne Paris Cité, Paris, France.
[4] Synchrotron SOLEIL, L'Orme des Merisiers, Départementale 128, 91190 Saint-Aubin, France.
[5] Centre de Nanosciences et de Nanotechnologies, CNRS, Université Paris-Saclay, C2N, Palaiseau 2110, France.
[6] New Imaging Technologies SA, 1 impasse de la Noisette 91370 Verrières le Buisson, France.
[7] Faculty of Physics and Applied Computer Science, AGH University, Al. Mickiewicza 30, PL-30-059, Kraków, Poland.
[8] Center for Nanoscale Materials, Argonne National Laboratory, 9700 S. Cass Avenue, Lemont, Illinois 60439, United States.
[9] Departament de Quimica Fisica i Analitica, Universitat Jaume I, E-12080, Castello de la Plana, Spain.

$ these two authors have equal contributions

*To whom correspondence should be sent: el@insp.upmc.fr, prado@insp.upmc.fr

**Abstract:** Transferring the nanocrystals (NCs) from the laboratory environment toward practical applications has raised new challenges. In the case of NCs for display and lightning, the focus was on reduced Auger recombination and maintaining luminescence at high temperatures. When it comes to infrared sensing, narrow band gap materials are required and HgTe appears as the most spectrally tunable platform. Its low-temperature synthesis reduces the growth energy cost yet also favors sintering. As a result, once coupled to a read-out circuit, the Joule effect aggregates the particles leading to a poorly defined optical edge and dramatically large dark current. Here, we demonstrate that CdS shells bring the expected thermal stability (no redshift upon annealing, reduced tendency to form amalgams and preservation of photoconduction after an atomic layer deposition process). The peculiar electronic structure of these confined particles is unveiled using k.p self-consistent simulations showing a significant exciton biding energy at around 200 meV. After shelling, the material displays a *p*-type behavior that favors the generation of photoconductive gain. The latter is then used to increase the external quantum efficiency of an infrared imager, which now reaches 40 % while presenting long-term stability.

In a few years' time scale, narrow band gap nanocrystals (NCs) have switched from the basic design of infrared absorbing optical transitions,[1] addressing questions such as the exact nature of their absorption[2] (*i.e.*, interband vs intraband[3,4]), to a spin-coatable semiconductor platform with reliable photoconductive properties. Thanks to this progress, the focus has evolved toward integrating these infrared NCs into complex devices far beyond the simple sandwiching of an absorbing layer between charge transporting/blocking layers. The coupling to photonic structures is one of the most striking developments achieved.[5,6] With such structures, light-matter coupling is no longer only driven by material properties but rather by the geometrical aspects of the cavity. Beyond boosting the infrared (IR) absorption magnitude,[7–10] such approaches have been relevant to generating new functionalities, including an actively reconfigurable photoresponse.[11,12] However, the most striking progress certainly relates to the transition from a single-pixel detector to a full infrared camera. PbS NC-based cameras have been demonstrated with extremely small pixel pitch[13–15] (*i.e.*, <2 µm whereas best strategies based on InGaAs still present pixel pitch larger than 5 µm). Moreover, HgTe focal plane array (FPA) operation[16,17] from near- to mid-IR[18,19] has been achieved with all possible operation modes: photoconductive,[18,20–22] photovoltaic[23] and even phototransistor.[24]

Despite containing heavy metal, these HgTe NCs have a reduced toxicologic impact with respect to thin films of HgCdTe by avoiding the need of epitaxial growth onto a CdZnTe wafer, the latter being main source of heavy metal in fabrication The environmental impact is also reduced by decreasing the energy consumption cost with material growth at low temperatures. In spite of the above developments, some central problems have been swept under the rug, particularly for device transition from single-pixels to imagers. As the number of pixels increases, more electrical current is required to drive the read-out integrated circuit resulting in more Joule heating. Such a factor is fundamentally problematic for infrared sensors in which there is always a competition for the activation of carrier between photons (*i.e.,* the signal) and phonons (*i.e.,* the noise). In the case of HgTe NCs, the problem appears to be even more dramatic due to the relatively low material thermal conductivity, which can easily result in sintering, even at modest heating. This process is illustrated in **Figure 1**c illustrates. Here, after being exposed to 100 °C for less than one hour, the photocurrent spectrum of a film presenting an initial 2 µm cut-off wavelength is redshifted above 5 µm. Far beyond the loss of optical sharpness, the drop of the band gap—which is not limited by bulk band gap, HgTe being a semimetal—leads to an exponential increase of carrier thermal activation and thus of dark current. This process has been identified as the main aging path for the HgTe NC-based focal plane arrays, more important than oxidation of the material.[25] Thus, strategies to increase the thermal stability of such NCs become critical for their effective integration into commercial products. Encapsulation of the NCs, for example using atomic layer deposition (ALD) may be seen as a strategy to prevent particle coalescence. However previous attempts to conduct ALD on HgTe NCs have also led to a complete drop of their photoconductive properties. NC shelling is also challenging since traditional methods to grow larger band gap semiconductors are generally conducted at high temperatures (200°C and more) where HgTe NCs are already damaged. This is why most reported attempts to shell mercury chalcogenide compounds actually rely on the so-called colloidal ALD method[26] conducted at room temperature.[2,27–29] Such shells have never been fully satisfactory with no striking increase of the luminescence quantum yield in the case of mercury chalcogenides. Alternatively, shelling at a reduced temperature can be obtained using a more reactive precursor. For perovskites, another class of fragile NCs, a monomolecular precursor has been proposed as a strategy to grow shells under mild conditions.[30] CdS shelling has also been recently investigated for HgSe using this approach,[31] but not for HgTe in which the lattice mismatch remains larger (≈10 %).

Here, we report the synthesis of HgTe-CdS core shell NCs with substantially enhanced thermal and chemical stability. Using k.p and self-consistent simulations, we analyze the material's electronic structure and provide an explanation for the limited spectral shift observed during the shell growth. We also stress how the peculiar band structure of bulk HgTe impacts the wavefunction of the material and also discuss how coulombic interactions might be larger than expected with binding energy as large as several hundred meV. We then probe the electronic transport properties that show a *p*-type behavior for the film after annealing that is favorable for photoconductive gain and thus large responsivity. At the single-pixel level, an external quantum efficiency (EQE) above 100 % can be obtained through this mechanism while the specific detectivity is estimated to be around $3\times10^{11}$ Jones at room temperature. We then further test the potential of this core-shell material for infrared imaging and demonstrate imaging with EQE now reaching 40 % (compared to 5 % for core-only NCs coupled to the same read-out circuit) and long integration times (>100 ms) thanks to reduced dark current.

We aim to develop a shell growth strategy for the HgTe cores that would improve their spectral stability upon high-temperature exposure. We aim for a thin shell since the latter will also act as tunnel barrier and will reinforce confinement which will be favorable for luminescence but detrimental for photoconduction. Furthermore, the large lattice parameter is unfavorable for large shell growth due to accumulation of strain.

Inspired by the recent work from Kamath *et al.* dedicated to the growth of CdS shells on HgSe cores,[31] we use cadmium bis(phenyldithiocarbamate) (Cd(PDTC)$_2$) as a monomolecular source of cadmium and sulfide. Our early attempts were performed using HgTe cores obtained using the Keuleyan's procedure.[32] This leads to sharp optical transitions in the short-wave infrared but also to a high degree of aggregation and a strong particle shape deviation from sphere[33] (Figure S1). CdS shell growth has indeed been observed but the particle appears strongly branched with long arms (10 nm and more, Figure S2). These branched-shaped particles appear highly unfavorable to obtain densely packed films, limiting light absorption and interparticle coupling, that are required for efficient charge conduction. This is why we later consider more spherically-shaped HgTe NCs obtained from a seeded approach.[34] As shown from electronic microscopy (**Figure 1**a) these core particles are not aggregated and have a narrow size distribution with an average size of 4.4 nm.

Compared to Kamath *et al*, the growth procedure has been modified and we found that more spherical-shaped shells are grown if the reaction is conducted in the presence of two cadmium sources (CdCl$_2$ is introduced in the reaction medium in addition to (Cd(PDTC)$_2$) ). The amount of precursor is chosen so that the shell remains as thin as one monolayer (ML). There are two reasons for that: first the 10 % lattice mismatch between HgTe and CdS is unfavorable to the growth of a high-quality thick shell. Secondly, a thick shell would be detrimental to transport given the wide band gap of CdS compared to HgTe. After shell growth, the particles appear slightly more faceted, see **Figure 1**b. After shelling, the absorption spectrum appears barely redshifted (25 meV, see Figure S3), asking for further confirmation of the shell growth. To avoid contribution from ligand excess and possible secondary nucleation, the particles have encountered several washing steps. The, both visible absorption (feature at 380 nm), Raman (presence of LO phonon mode from CdS), X-ray diffraction (diffraction peak matching the (111) plane of CdS which is also the most intense peak of CdS in the zinc blende phase) and X-ray photoemission spectroscopy (presence of Cd 3d states) confirm the material presence (Figure S3-6). The most striking evidence for the shell growth comes from the lack of spectral shift upon annealing that would occur if there were sintering, see **Figure 1**d.

We have determined the thermal conductivity (Figure S7) of both core (0.9 ± 0.2 W/m•K) and core-shell (1.0 ± 0.2 W/m•K) materials in films and found similar magnitude for the two of them. Thus, higher stability does not result from the better capacity of the core-shell to extract heat but rather from an inherent stabilization by the CdS shell. The core-shell is better suited for withstanding passive storage under elevated temperatures, such as those encountered during annealing in lithography and encapsulation processes. This enhances its potential for facilitating later device integration. Another benefit of the shell is to suppress the tendency of HgTe NCs to form an amalgam with metal,[35] especially silver. Indeed, the core-shell NCs coupled to an Ag film do not show any transformation of the HgTe into an Hg-Ag amalgam (Figure S5).

The material is also luminescent (**Figure 1**e) and the PL decay appears mono-exponential as a function of time (**Figure 1**f) with a decay time at around 45 ns, which may be an indication of suppressed non-radiative pathways compared to more aggregated NCs for which the decay curve presents multiple decay constants spanning from the ns range to several hundreds of ns.[36–39]

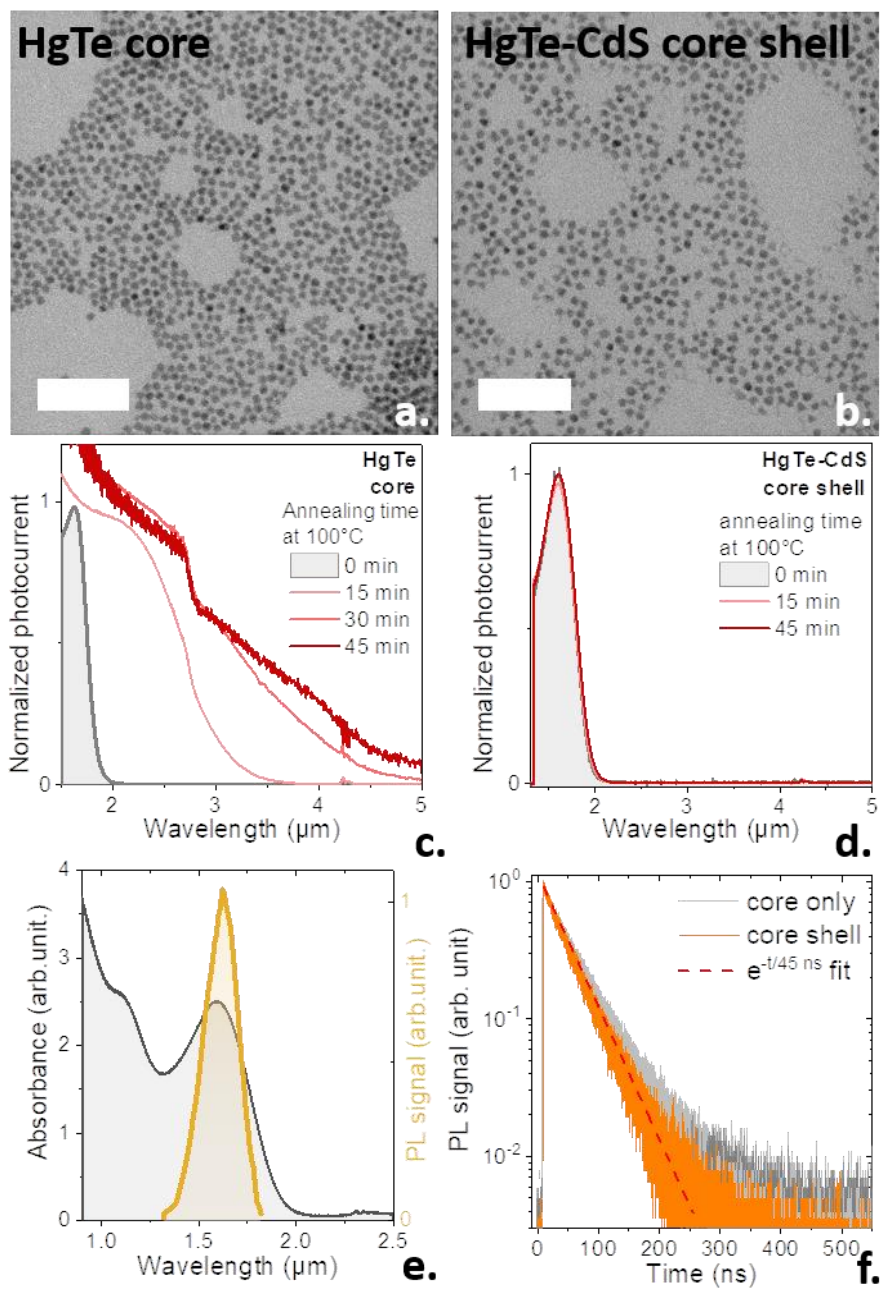

*Figure 1 Core shell HgTe-CdS nanocrystals with high thermal stability. a. TEM image of non-aggregated quasi sphere shape HgTe core used to later obtaine core shell. b. TEM image of HgTe-CdS core-shell. In part a and b scale bars are 50 nm long. c. (resp d.) Photocurrent spectra from a thin film of HgTe cores (resp HgTe-CdS core-shell) before and after annealing at 100 °C and for various durations. e. Absorption and photoluminescence spectra for the HgTe-CdS core-shell film. f. Time resolved photoluminescence from the HgTe core and HgTe-CdS core-shell measured in solution under pulsed excitation at 705 nm.*

To unveil the electronic structure of such core-shell particles and determine their specific properties, we used 8 band 3D k.p modelling coupled with self-consistent simulations (see Figure S8 and associated discussion). Due to the large hole effective mass in HgTe, the density of states in the valence band is quite high with a *p*-like ground state ($1P_{3/2}$), whereas the *s*-like state ($1S_{3/2}$) appears at an energy difference (31 meV) similar to thermal energy at room temperature and will thus be partly overlapping with band edge transition, see **Figure 2**a. In HgTe, the narrow band gap leads to a strong mixing of $\Gamma_6$-$\Gamma_8$ bands. This makes *s* orbitals of the conduction band couple to *p* orbitals of the valence band. Such a *s/p* mixing results in the electron state being more spread toward the surface than a purely *s* state, see **Figure 2**b. This behavior is further illustrated by looking at the radial distribution of the state charge density, which is pushed toward the surface (green line) compared to the case where mixing is artificially suppressed using a positive value for the HgTe band gap (orange line). Such behavior is interesting since it means that the shell will not prevent the electron wavefunction from reaching the surface, which is critical for interparticle coupling that drives the electronic transport magnitude.

It is also worth noticing that the strong dielectric mismatch heavily impacts the confinement magnitude, see **Figure 2**c. Though HgTe presents a relatively large dielectric constant ($\varepsilon_{HgTe}>20$),[40] the combination of small size and small external dielectric constant resulting from ligand and void in the NC film makes the binding energy quite large (≈200 meV for $\varepsilon_{out}=3$,[40] see **Figure 2**d), compared to the band gap (≈700 meV). Such a large binding energy is an unanticipated consequence of the effort to use more spherically shaped particles. This was motivated to favor shell growth and particle packing (to obtain large effective dielectric constant and increase effective mobility), but this makes charge dissociation more difficult compared to branched shape particles[33].

However, we do not observe a strong excitonic signature in the spectrum because the electron-hole interaction is mostly balanced by the self-energy term, see **Figure 2**e. The monolayer shell growth only induces a very limited spectral shift (-25 meV). This situation contrasts with Cd-based NCs for which a shell growth is associated with a much stronger redshift.[41] There, the shell growth leads to a more favorable delocalization of the wavefunction, toward the surface compared to vacuum/ligand surrounding due to the reduced band offset between material. A plausible explanation is that the resulting redshift is balanced by an effect of pressure. As discussed in Figure S9 and table S1, the 10 % mismatch between the core and the shell lattice is expected to induce a 1.7 GPa pressure on the core.[42] Such a pressure is larger than the threshold pressure (1.4 GPa) for the zinc blende to cinnabar transition[43,44] in bulk HgTe. However, for NCs, surface tension tends to stabilize the low-pressure phases for a wider pressure range. Nevertheless, pressure induces a +60 meV/GPa blue shift of the exciton,[45] which balances the weaker confinement resulting from shell growth. In sum, pressure and confinement terms mostly cancel each other out. Whereas electrostatic renormalization of the band gap remains weak, exciton binding energy for this strongly confined form of HgTe appears as a significant part of the band gap.

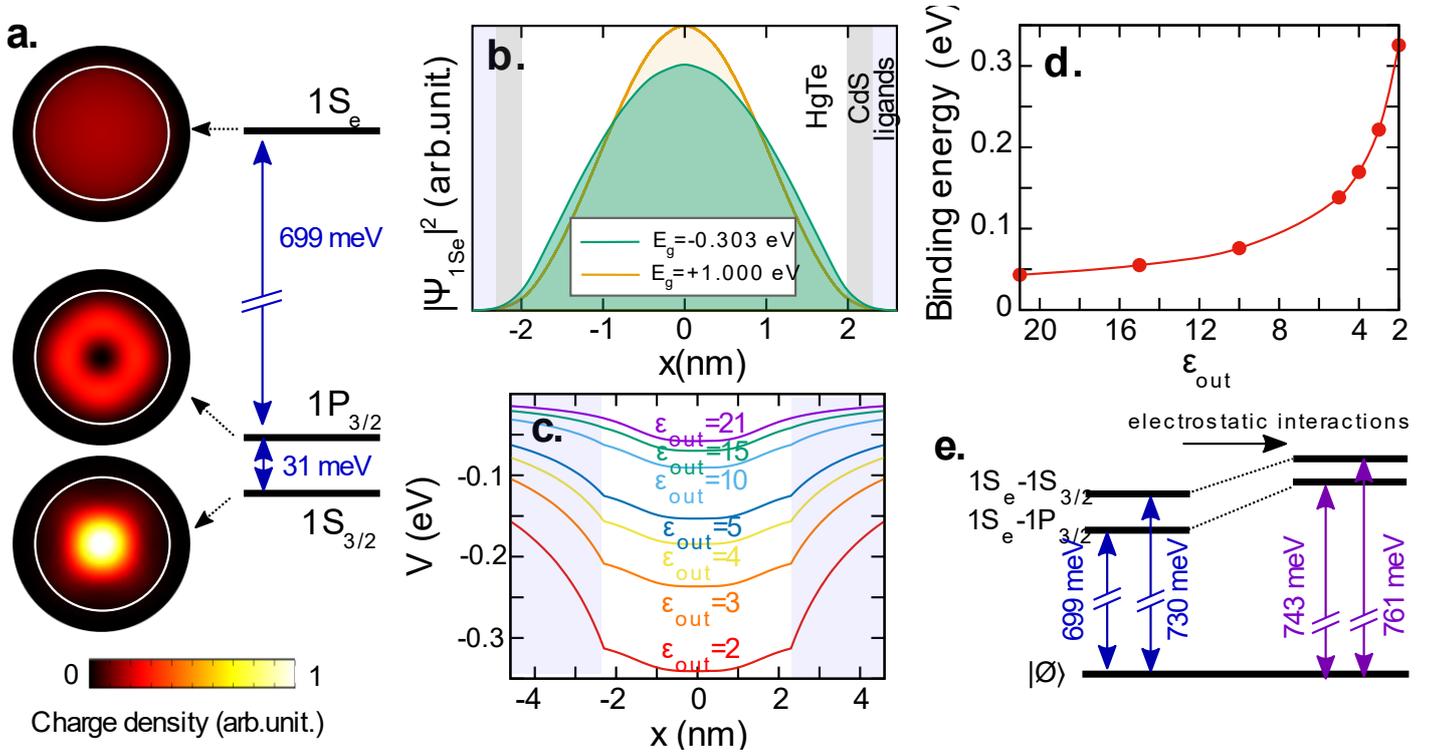

*Figure 2 Modelling core shell electronic structure.* a. Effective electronic states in the vicinity of the Fermi level. For each state we also plot the associated charge density. The white lines depict the transition from HgTe core to CdS shell. b. Probability of finding the electron in the plane orthogonal to the cross-section coordinate x, evaluated for the bulk $\Gamma_8$-$\Gamma_6$ splitting ($E_G$=-0.303 eV) and for an artificial value ($E_G$=+1 eV) which quenches the s/p mixing in the conduction band. c. Electrostatic potential generated by the 1 $S_e$ for various values of the surrounding dielectric constant. d. Binding energy for the exciton as a function of the surrounding dielectric constant. e. Ground and first excited transition without and with electrostatic interactions (self energy and electron-hole Coulomb attraction).

In a third step, we investigated the conductive and photoconductive properties of HgTe-CdS NCs. Since, we found the material less conductive than the pristine cores, we annealed the films under mild condition in air, raising the overall film conductivity (**Figure 3**a) without inducing a spectral shift of the response. However, we note that the transistor characteristic of the film gets affected by the annealing step, see **Figure 3**b-c. Before annealing the material presents an ambipolar behavior in which the transfer curves have both a hole and an electron branch, with the neutrality point occurring under positive gate bias (*i.e.,* the film is initially p-type). After annealing only hole conduction is observed. Since the band edge energy remains unchanged (**Figure 1**d), we conclude that the rise of conduction is due to a shift of the Fermi level toward the valence band, increasing the hole density. The stronger *p*-character favors the generation of the photogating effect. Field effect transistor measurements (Figure S10) also reveals an overall reduced (factor 100) carrier mobility after the shell growth. Again, this situation contrasts with Cd based NCs for which spectral shift is large but mobility remains mostly un affected by shell growth. With narrow band gap material, such as HgTe, the shell immediately behave as a tunnel barrier inducing the drop of mobility. This observation further pledges for the growth of thin shell only, but also found a benefit for imaging since reduced mobility would also generate less interpixel cross talk.

Under illumination, both electrons and holes get generated, however the larger hole mobility favors its transportation while the electron will stay in the core-shell acting as traps. Once a hole reached the electrode, the latter will be reinjected by the opposite electrode to ensure the sample neutrality, it becomes possible to generate several carriers per absorbed photon and to obtain an external quantum efficiency (EQE) larger than unity. We then probe the photoconductive properties using a geometry that matches the one used in the following for the imaging setup (*i.e.,* square metallic pads of 6 µm that are organized within a square array of 15 µm period). As a result of photogating, fairly large responsivities reaching 1.2 A.W$^{-1}$ under 3 V of bias (**Figure 3**e) have been obtained. The noise spectral density, though limited by 1/f noise, (**Figure 3**f) enables large detectivity reaching 3x10$^{11}$ Jones at room temperature thanks to the photoconductive gain.

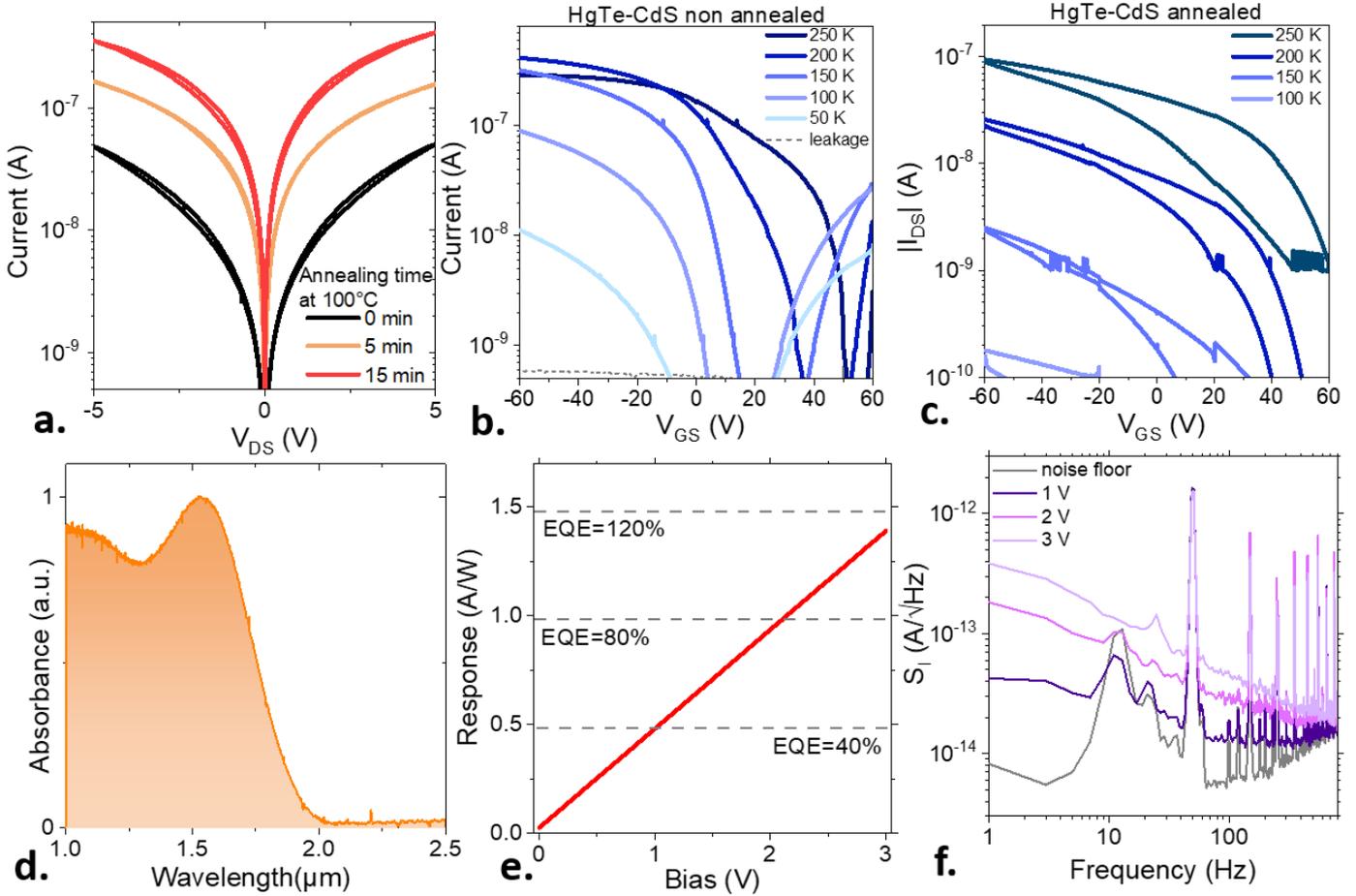

*Figure 3 Transport and photoconduction in HgTe-CdS NCs array. a. IV curves for an array of HgTe-CdS core-shell NCs after annealing for various durations. b. Transfer curves (drain current vs gate voltage under constant drain bias) at various temperatures for HgTe-CdS core-shell NCs. c. Transfer curves at various temperatures for HgTe-CdS core-shell NCs after annealing at 70°C for 1h. d. Photocurrent spectrum for HgTe-CdS core-shell NCs after annealing at 70°C for 1h. e. Responsivity for thin film of HgTe-CdS core-shell NC film as a function of the applied bias. f. Noise spectral current densities from HgTe-CdS core-shell NC film under various biases. Parts d to f are measured at room temperature in air.*

Now that we have established the photoconductive character of the HgTe-CdS material, we test its potential for imaging. The NC solution is transformed into an ink and is deposited onto a read-out integrated circuit (ROIC). The latter has a VGA format (640x512 pixels, with a 15 µm period). Compared with our previous generation of ROIC,[21] the surface has been planarized using a polishing step and by growing metallic pixel electrodes whose height matches the surrounding dielectric.[23] Such a planar surface reduces crack formation during the ink spin-coating. The ROIC (NITquantum 1601C) also enables to apply a checkerboard bias distribution and, thus, an in-plane electric field. As it is, this approach sacrifices half the resolution since one pixel is used to apply bias and the other one to collect charges. However, the high frame rate of our ROIC makes that the checkerboard distribution can be alternated one frame out of two and we are now able to acquire full frame VGA format images.

A series of images acquired this way is shown in **Figure 4**a-d and Figure S11-15. Compared to the imager based on pristine cores, the lower conductivity of the film reduces dark current and thus enables a longer integration time a few ms at room T and up to 200 ms at -40°C, values that are comparable to InGaAs. Moreover, the photoconductive gain mechanism is responsible for improved EQE that was only 4-5 % for core only and now reaches 40 %, see **Figure 4**e-f. A comparison of the imager performances of the core and core/shell is proposed in Table 1, highlighting that the current device is showing the best trade-off between performances and image quality for devices relying on HgTe NCs. Regarding performance metrics, overcoming the poor stability of HgTe devices was a target of this work. In the previous device based on core only (Figure S14) where thermal stability is an issue including for simple passive storage, a dramatic drop of the image quality was observed after 24 h of continuous operation with an increase in the dark current by a factor of 2 over this time scale. With the core-shell NCs, the dark current only changes by a few % after 2880 h of air storage (to test chemical stability) and after 48h of continuous operation (stability toward Joule effect) demonstrating long-term use of core-shell-based devices, as opposed to the limited lifetime of their core-only-based counterpart. Another striking benefit of the shelling is the improved stability of the material toward traditional semiconductor processing. To illustrate that we have conducted an ALD process to encapsulate the film with an $Al_2O_3$ layer (5 nm deposited over 5h at 50°C, see Figure S15). Previous attempts to perform such encapsulation of HgTe NCs using ALD led to an increased film conductivity, but a complete loss of photoconductivity.[46] With the shelled particles, the material maintains its conductance and photoconductance and its photoconductive spectrum remains mostly unaffected (Figure S15).

*Table 1 Imaging setup based on HgTe NCs. PC stands for photoconductor, PT for phototransistor and PV for photovoltaic. RT is used for room temperature*

| Operating mode | Format and pixel size | Cut off wavelength (µm) | Operating temperature | Dark current | EQE (%) | Detectivity (Jones) | References |
|---|---|---|---|---|---|---|---|
| PC | 320x256<br>30 µm | 1.5 (SWIR)<br>5 (MWIR) | RT (SWIR)<br>80 K MWIR | | 175 | 2x10$^{11}$ (SWIR)<br>8x10$^{10}$ (MWIR) | [18] |
| PC | 320x256<br>30 µm | 5 | 120 K | 4 pA | 0.75 | | [19] |
| PV | Full VGA – 15 µm | 2 | -30°C | | 0.3 | | [23] |
| PC | 1280x1024 -15 µm | 2 | RT | | 14 | 2.8x10$^{10}$ | [20] |
| PC | Half VGA | 1.8 | -30°C | | 5 | - | [21] |

| | | | | | |
|---|---|---|---|---|---|
| PC | Full VGA – 15 µm | 2 | -40°C to RT | 40 | **This work** |

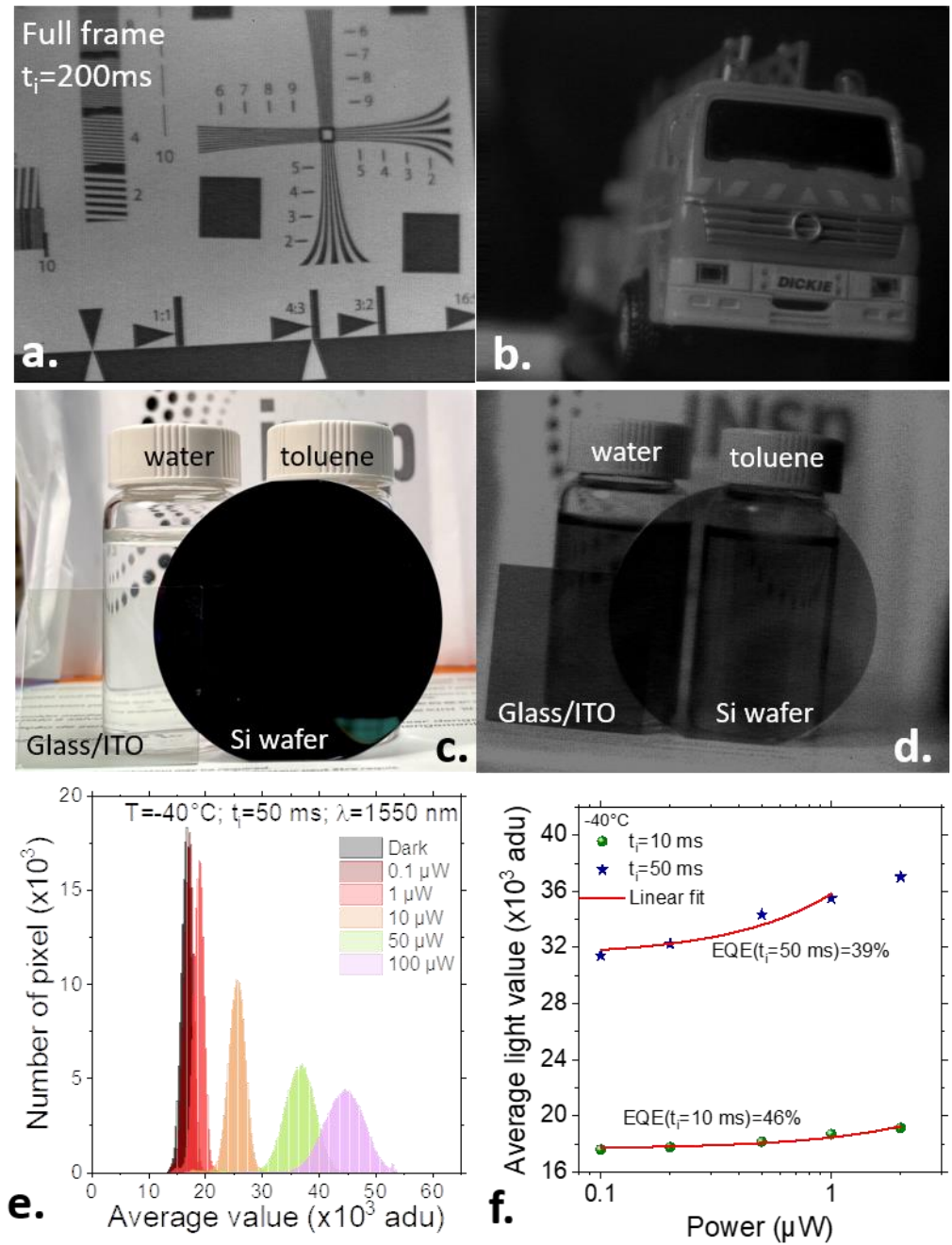

*Figure 4 Infrared imaging using HgTe-CdS NCs as absorbing layer. a. Full frame VGA format image from a test chart, acquired with a 200 ms acquisition time. b. Half VGA format image from a toy truck, acquired with a 200 ms integration time ($t_i$). c. Visible image from a scene containing a glass coated ITO wafer (transparent in the visible), a silicon wafer (reflective in the visible) and two vials containing transparent liquids (water and toluene). d. Infrared image of the same scene acquired with the NC-based camera. Parts a, b and d are acquired at -40 °C. e. Dark and under illumination signal amplitude of the pixel under various incident powers. f. Pixel signal magnitude as a function of the incident illumination power for two integration times. The slopes of this graph direcly relate to the external quantum efficiency.*

To summarize, we have demonstrated the CdS shelling of HgTe NCs to obtain thermally and chemically stable narrow band gap NCs with absorption in the short-wave infrared. The thin shell (1 ML) appears compatible with transport and above all suppresses thermal degradation that usually induced spectral shifts of the material upon mild annealing. The shell growth only induces a marginal spectral shift since reduced confinement is mostly balanced by the pressure effect. Similarly, regarding the electrostatic correction, the electron-hole interaction and self-energy also cancel each other out in spite of the large exciton binding energy. This is a problem that has certainly been underestimated in the past and that future material developments would have to address. Thanks to reinforced *p*-character, the shelling also favors the formation of a *p*-type film with photoconductive gain that enables large responsivity (>1 A.W$^{-1}$) and detectivity (3x10$^{11}$ Jones at room temperature). Once coupled to a read-out circuit, the core-shell material presents reduced dark current compared to pristine-only material, enabling longer integration times for the sensor operation. The photoconductive gain enables improved EQE (now reaching 40 % as opposed to 5 % for core only material). Moreover, the stability of the device has been extensively improved, enabling for the first-time long-term operation at the focal plane array level.

## SUPPORTING INFORMATION

Supporting Information include details about (*i*) Methods, (ii) Morphology of the core-shell as a function of core growth and shell growth procedure, (*iii*) Material characterization of core-shell material, (*iv*) thermal conductivity determination, (*v*) Electronic structure modelling using k.p and self-consistent methods, (*vi*) Effect of shell induced pressure on the core, (*vii*) carrier mobility and (*viii*) Performances and aging of the core shell camera.

## ACKNOWLEDEGMENTS

We thank Xiang Zhen Xu for performing TEM imaging and Michael Rosticher for performing ALD deposition. The project is supported by ERC grants blackQD (grant n° 756225), Ne2dem (grant 853049), and AQDtive (grant n°101086358). This work was supported by French state funds managed by the Agence Nationale de la recherche (ANR) through the grant Copin (ANR-19-CE24-0022), Frontal (ANR-19-CE09-0017), Graskop (ANR-19-CE09-0026), NITQuantum (ANR-20-ASTR-0008), Bright (ANR-21-CE24-0012), MixDFerro (ANR-21-CE09-0029), Quicktera (ANR-22-CE09-0018), Operatwist (ANR-22-CE09-0037) and E-map (ANR-22-CE50). Work performed at the Center for Nanoscale Materials, a U.S. Department of Energy Office of Science User Facility, was supported by the U.S. DOE, Office of Basic Energy Sciences, under Contract No. DE-AC02-06CH11357. H.Z. thanks the Chinese Scholarship Council for PhD funding. BS acknowledges support from the program "Excellence initiative – research university" for the AGH University of Krakow. J.I.C acknowledges support from Grant PID2021-128659NBI00, funded by MCIN/AEI/10.13039/501100011033 and "ERDF A way of making Europe".

## AUTHOR CONTRIBUTIONS

H.Z. and Y.P. developed the core and core-shell growth procedure. H.L. and B.D. conducted PL and time resolved PL measurements. D.F. and J.U. conducted thermal conductivity measurements. A.M. and D.P. conducted Raman measurements. C.D. and S.I. conducted X-ray diffraction. H.Z., R.A., M.C., T.D., A.K., E.B. conducted XPS measurements under supervision of D.P. using MS beamline. H.Z., N.L. and R.A. performed the transport measurements. B.S. and J.C. performed electronic structure simulations. Read-out circuit are provided and optimized by V.P. and P.P. Imaging in the infrared is conducted by R.A. and A.K. using a setup developed by D.D. E.L. makes project funded and wrote the manuscript with inputs from all authors.

**CONFLICT OF INTEREST**
The authors declare no competing financial interest.

**TOC graphic**

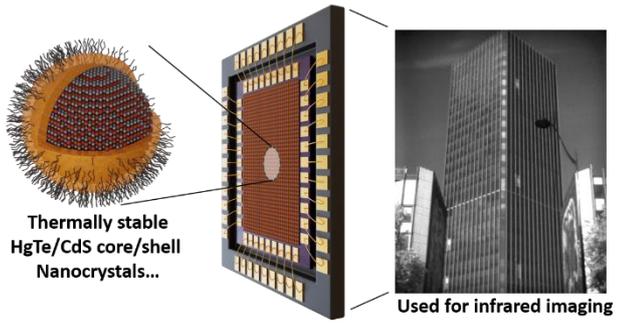

*Supporting information for*

**Infrared Imaging using thermally stable HgTe/CdS nanocrystals**

Huichen Zhang[1$], Yoann Prado[1$,*], Rodolphe Alchaar[1], Henri Lehouelleur[2], Mariarosa Cavallo[1], Tung Huu Dang[1,3], Adrien Khalili[1], Erwan Bossavit[1,4], Corentin Dabard[2], Nicolas Ledos[1], Mathieu G Silly[4], Ali Madouri[5], Daniele Fournier[1], James K. Utterback[1], Debora Pierucci[1], Victor Parahyba[6], Pierre Potet[6], David Darson[3], Sandrine Ithurria[2], Bartłomiej Szafran[7], Benjamin T. Diroll[8], Juan I. Climente[9], Emmanuel Lhuillier[1*]

[1] Sorbonne Université, CNRS, Institut des NanoSciences de Paris, INSP, F-75005 Paris, France.
[2] Laboratoire de Physique et d'Etude des Matériaux, ESPCI-Paris, PSL Research University, Sorbonne Université, CNRS UMR 8213, 10 rue Vauquelin 75005 Paris, France.
[3] Laboratoire de Physique de l'Ecole normale supérieure, ENS, Université PSL, CNRS, Sorbonne Université, Sorbonne Paris Cité, Paris, France.
[4] Synchrotron SOLEIL, L'Orme des Merisiers, Départementale 128, 91190 Saint-Aubin, France.
[5] Centre de Nanosciences et de Nanotechnologies, CNRS, Université Paris-Saclay, C2N, Palaiseau 2110, France.
[6] New Imaging Technologies SA, 1 impasse de la Noisette 91370 Verrières le Buisson, France.
[7] Faculty of Physics and Applied Computer Science, AGH University, Al. Mickiewicza 30, PL-30-059, Kraków, Poland.
[8] Center for Nanoscale Materials, Argonne National Laboratory, 9700 S. Cass Avenue, Lemont, Illinois 60439, United States.
[9] Departament de Quimica Fisica i Analitica, Universitat Jaume I, E-12080, Castello de la Plana, Spain.

$ these two authors have equal contributions

*To whom correspondence should be sent: el@insp.upmc.fr, prado@insp.upmc.fr

# Table of content

## 1. METHODS

**Chemicals:** Mercury chloride (HgCl$_2$, Sigma Aldrich, 99%), mercury bromide (HgBr$_2$, Alfa Aesar), mercury iodide (HgI$_2$, Touzart & Matignon), Cadmium chloride (CdCl$_2$, Alfa Aesar, 99.99%), tellurium powder (Te, Alfa Aesar, 99.99%), trioctylphosphine (TOP, thermofisher scientific, 90%), oleylamine (OLA, Acros, 80-90%), octadecene (ODE, Acros Organics, 90%), dodecanethiol (DDT, Sigma-Aldrich, 98%), oleic acid (OA, Alfa Aesar, 90%), methanol (MEOH, Carlo Erba, 99.8%), acetone (VWR), absolute ethanol (EtOH, VWR), isopropanol (IPA, VWR), toluene (Carlo Erba, 99.3%), chlorobenzene (VWR), N,N dimethylformamide (DMF, VWR), 2-mercaptoethanol (MPOH, Merck, >99%), methylisobutylketone (MIBK, VWR, >98.5%), 3-mercaptopropionic acid (MPA, Alfa Aesar, 99%). All chemicals are used as received, except oleylamine which is centrifuged before used. **Mercury compounds are highly toxic. Handle them with special care.**

**Ammonium phenyldithiocarbamate synthesis (NH$_4$PDTC)**[1,2]**:** In a 3-neck 100 mL flask under nitrogen bubbling at room temperature, equipped with a trap of NaClO, 25 mL of NH$_4$OH, 25 mL acetone and aniline (10 mL, 0.11 mol) were introduced and cooled down with an ice bath. Then, CS$_2$ (10 mL, 0.16 mol) was added dropwise with vigorous stirring. The solution turned red after 5 minutes, and a pale-yellow crystalline product slowly precipitated over the following 90 minutes. The product was collected by vacuum filtration and washed with 4 times CHCl$_3$. The NH$_4$PDTC was dried under vacuum and stored in a freezer to prevent decomposition.

**Cadmium bis(phenyldithiocarbamate) synthesis [Cd(PDTC)$_2$]**[1] **:** In an Erlenmeyer flask, NH$_4$PDTC (600 mg, 3.2 mmol) was dissolved in 50 mL water. The solution was clear yellow. Then, Cd(NO$_3$)$_2$.4H$_2$O (493 mg, 1.6 mmol) was dissolved in 40 mL water and added dropwise to the flask over 5 minutes, leading to immediate precipitation of Cd(PDTC)$_2$ as a pale yellow powder. The product was collected by centrifugation, and washed with 25 mL of absolute ethanol. The precipitate was ground with a mortar and pestle, dried under a Schlenk line and stored in a desiccator. Yield: (Mw=449 g.mol$^{-1}$)

**1 M TOP:Te precursor**: 6.35 g of Te powder was mixed in 50 mL of TOP in a three-neck flask. The flask was kept under vacuum at room temperature for 5 min and then the temperature was raised to 100 °C. Furthermore, degassing of flask was conducted for the next 20 min. The atmosphere was switched to Ar and the temperature was raised to 275 °C. The solution was stirred until a clear orange coloration was obtained. The flask was cooled down to room temperature and the color switched to yellow. Finally, this solution was transferred to a nitrogen-filled glove box for storage.

**HgTe core with 5000 cm$^{-1}$ (2 µm) cut-off:** In a 50 mL three-neck flask, 18 mL of oleylamine was degassed under vacuum at 110°C for 1 h. Then the atmosphere was switched to N$_2$ and the temperature was set at 90°C. Meanwhile, in a 20 mL vial, 144 mg (0.4 mmol) of HgBr$_2$ was dissolved in 3 mL of oleylamine under sonication and degassed at 110°C for 30 min. Then, the solution in the vial was cooled down to room temperature and switched to N$_2$ atmosphere. 0.2 mL of TOP:Te (1 M) with 36 µL (0.15 mmol) of dodecanethiol (DDT) was injected into the vial. The solution turned orange quickly and was injected into the hot oleylamine in flask through a syringe. After 30 s, the reaction was quenched with 1 mL of a solution of DDT in toluene (10% v/v) and ice bath. The content of the flask was transferred to a falcon and precipitated by adding MeOH. The pellet was redispersed in toluene. The nanocrystals were precipitated a second time with absolute EtOH and redispersed in toluene. The toluene solution was centrifugated to remove the unstable phase and then filtered with 0.2 µm PTFE filter. This material is the one used for the detection side.

**HgTe core with 7000 cm$^{-1}$ (1.45 µm) band edge:** In a 50 mL three-neck flask, 18 mL of oleylamine was degassed under vacuum at 110°C for 1 h. Then the atmosphere was switched to N$_2$ and the temperature was set at 120°C. Meanwhile, in a 20 mL vial, 144 mg (0.4 mmol) of HgBr$_2$ was dissolved in 5 mL of oleylamine under sonication and degassed at 110°C for 30 min. Then, the solution in vial was cooled down to room temperature and switched to N$_2$ atmosphere. 0.3 mL of TOP:Te (1 M) with 20 µL (0.23 mmol) of 3-mercaptopropanoic acid (MPA) was injected into the vial. The solution was injected into the hot oleylamine in the flask through a syringe. After 1 min, the reaction was quenched with 1 mL of a solution of DDT in toluene (10% v/v) and ice bath. The flask content was transferred to a falcon and was precipitated by adding MeOH and centrifugation. The pellet was redispersed in toluene. The nanocrystals were precipitated a second time with absolute EtOH and redispersed in toluene. The toluene solution was centrifugated to remove the unstable phase and then filtered with 0.2 µm PTFE filter. This material is used for PL to ensure that the signal matches our time resolved detection capability.

**Branched shell growth:** 12 mL of OLA was degassed in a 50 mL three-neck flask at 110°C for 1h before being cooled down to room temperature and switched to $N_2$ atmosphere. 0.76 mL of HgTe core solution (1.15 OD at 400 nm after a 500x dilution) was introduced into the flask and 3 cycles of vacuum/$N_2$ were performed. The temperature was then set at 80°C, and $Cd(PDTC)_2$ (45 mg, 0.87 mmol) dissolved in 3 ml of a 20% (v/v) oleylamine-octadecene solution was injected when the temperature reached 40°C. The solution was kept at 80°C for 2 min and then quenched with 1 mL of 10% DDT in toluene solution and ice bath. The content in the flask was transferred to a centrifuge tube and precipitated by adding MeOH. After being redispersed in toluene, the nanocrystals were precipitated again with EtOH and redispersed in toluene. After redispersion, the toluene solution was centrifugated to remove the unstable phase (the supernatant was kept).

**Spherical shell growth:** 18 mg $CdCl_2$ was added to 18 mL of a 50% (v/v) oleylamine-octadecene solution in a flask and was degassed at 110°C for 30 min. The mixture was then cooled down to 40°C and switched to $N_2$ atmosphere. 3.5 mL of HgTe core solution (0.45 OD at 400 nm after a 500x dilution) was introduced to the flask, and 3 cycles of vacuum/$N_2$ cycles were conducted. $Cd(PDTC)_2$ (50 mg, 1 mmol) in 1.5 mL of a 30% (v/v) oleylamine-octadecene mixture was then injected at 40°C quickly, and the flask was heated up to 80°C for 2 min. The reaction was quenched by injecting 2 mL of a 10% (v/v) DDT/OA-toluene solution and cooling down to room temperature. The content in the flask was transferred to a centrifuge tube and precipitated with EtOH. The pellet was redispersed in toluene and then precipitated again with IPA. After redispersion, the toluene solution was centrifugated to remove the unstable phase (the supernatant was kept).

**HgTe/CdS ink*:*** 3 mL of the HgTe-CdS core-shell solution in toluene (0.9 OD at 400 nm after a 500x dilution) is mixed with 2 mL of exchange solution (15 mg $HgCl_2$, 9 mL DMF, 1 mL MPOH). The solution is stirred throughout vortex and sonication for 3 min to help the ligand exchange. Then, the solution is precipitated by adding toluene and centrifuged at 6000 rpm for 3 min. The supernatant is discarded and the QDs are dried under vacuum. Finally, the NCs are redispersed in 250 µL of DMF, centrifuged for 2 min at 6000 rpm and filtered through a 0.22 µm PTFE filter.

*Transmission electron microscopy:* For TEM imaging, a drop of diluted NCs solution is casted on a copper grid covered with an amorphous carbon film. The grid is degassed overnight under secondary vacuum. A JEOL 2010F is used at 200 kV for the acquisition of pictures.

**Infrared absorption spectroscopy:** A Fourier transform infrared spectrometer (FTIR) Fischer Nicolet iS50 in attenuated total reflection (ATR) configuration is used. The source is a white halogen lamp, the beamsplitter is made of $CaF_2$ and the detector is a DTGS ATR. Spectra are typically acquired with a 4 $cm^{-1}$ resolution and averaged over 32 spectra.

**Visible absorption spectroscopy:** spectra are acquired using solution of NC within a glass cuvette and measured using a Jasco 730 double path spectrometer with a 1 nm resolution.

**Infrared photoluminescence spectroscopy**: A vial containing a solution of NCs is illuminated bua bleu flash light. This light is used as a source of an FTIR (Thermo Scientific Nicolet iS50). The beamsplitter is made of $CaF_2$ and the signal collected using an extended InGaAs detector (2.6 µm cut-off uncooled). The spectra are generally averaged 256 times and acquired with a 4 $cm^{-1}$ resolution.

**X-ray diffraction (XRD)** signals from a film of nanoparticles drop-cast on a silicon substrate are recorded on a Phillips X'Pert diffractometer with a Cu-Kα radiation. Measurements are performed at a working condition of 40 kV in voltage and 40 mA in current.

**Energy dispersive X-ray spectroscopy:** The nanocrystals were deposited onto a Si substrate by drop-casting to make a ≈1 µm thick film. The SEM images were acquired using a Zeiss Supra 40 scanning electron microscope. The EDX acquisition was made in an area of 10 µm x 10 µm using a Brucker Quantax EDX XFlash 6/30. The acceleration bias was set to 10 kV and the aperture at 30 µm.

**Raman Spectroscopy:** the measurements were performed at room temperature using a commercial Renishaw inVia confocal Raman microscope with a 100x objective and a 532 nm laser excitation. The spectrometer calibration was set using the 520 ±0.5 $cm^{-1}$ band of a Si (100) single crystal. Spectra were recorded between 80 and 380 $cm^{-1}$ with 2400 lines/mm grating resulting in a resolution of 0.8 $cm^{-1}$. The effective laser power at the exit of the objective was about 0.07 mW.

**Time resolved photoluminescence measurements:** Time-resolved photoluminescence measurements were performed using a 705 nm laser diode (Picoquant) pulsed at 1 MHz, with emission collected by an optical fiber. The emission was directed through a monochromator to select the wavelength of interest and detection was performed using a superconducting nanowire array.

**High-resolution photoemission spectroscopy:** The measurements were performed at Tempo beamline[3] of synchrotron Soleil. Films of NCs were spin-cast onto a Si substrate coated with 80 nm of gold. The ligands of the NCs were exchanged using the same procedure as for device fabrication to avoid any charging effects during measurements. The samples were introduced in the preparation chamber and degassed under a vacuum below $10^{-9}$ mbar for at least two hours. Then the samples were introduced into the analysis chamber. The signal was acquired with an MBS A-1 photoelectron analyzer. The acquisition was done at a constant pass energy (50 eV) within the detector using a photon energy of 700 eV. A gold substrate was used to calibrate the Fermi energy.

**Electrodes fabrication for transport and photoconduction at single pixel level:** A polished $Si/SiO_2$ substrate was first rinsed and then sonicated in acetone for 5 min. The wafer was then rinsed with acetone and isopropanol, dried with a $N_2$ flow and finally cleaned with an $O_2$ plasma for 5 min. Then, AZ 5214E resist was spin coated on the substrate and baked at 110 °C for 90 s. The resist was exposed through a shadow mask to UV illumination for 1.5 s. The resist was baked again at 125 °C for 2 min and re-exposed to UV (without the mask) for 40 s. The resist was then developed using AZ 726 for 30 s. The film was rinsed with water and dried. The substrates were cleaned with an $O_2$ plasma for 5 min before 5 nm of chromium and 80 nm of gold were evaporated. The remaining resist was removed by immersing the film in acetone for one hour. The electrodes were then rinsed with isopropanol and finally dried. For field-effect transistor measurements, we used an interdigitated electrodes configuration that include 25 pairs of digits, each 2.5 mm long with a 20-µm gap between them. For the photoconduction response, we use a configuration closer to the one involved in the read-out integrated circuit with 2 square pads of 6 µm size with a period of 15 µm.

**Field-effect transistor characterization:** HgTe ink and HgTe/CdS ink were diluted by a factor 2 before being spin-coated on the interdigitated electrodes at 1800 rpm for 60 s to form a film of thickness ≈ 80 nm. The sample is mounted on the cold finger of a closed cycle He cryostat. The measurements have been conducted at 50, 100, 150, 200 and 250 K under secondary vacuum. The sample is connected to a Keithley 2634b which sets the drain bias ($V_{DS}$), controls the gate bias ($V_{GS}$) with a step between 50 and 200 mV and measures the associated currents $I_{DS}$ and $I_{GS}$. The mention "annealed" means that the film has been heated to 70°C, directly on a hot plate, for 15 minutes under inert atmosphere.

**Optoelectronic characterization**: Photoconductive single pixel devices were characterized at room temperature. For responsivity measurements, the samples were illuminated by an Omega BB-4A blackbody operated at 980°C with a 1 µm high pass filter, modulated by an optical chopper at 100 Hz. The signal was acquired with a Rohde & Schwarz RTE 1102 oscilloscope. The total power calculated according to the formula: $P = A_D . \pi . cos(\beta) . sin^2(\alpha) . \int_{\lambda_{min}}^{\lambda_{max}} \frac{hc^2}{\lambda^5} . \frac{1}{e^{hc/\lambda kT}} d\lambda$ where α is the solid angle illuminated, β is the angle of the sample (0° corresponds to sample perpendicular to the light illumination), $A_D$ is the device area, h is the Planck constant, c the light velocity, k is the Boltzmann constant and T the temperature. Noise measurements were acquired using a Stanford research systems model SR780 signal analyzer. The photocurrent spectra were measured with a Thermo Scientific Nicolet iS50 FTIR, between 10000 and 2000 cm$^{-1}$ averaged over 64 spectra with a 4 cm$^{-1}$ resolution. For all optoelectronic measurements, bias was applied and the signal was amplified by a FEMTO DLPCA-200 amplifier.

**Image sensor and coupling to a read-out circuit:** A read-out integrated circuit (ROIC – model NITquantum 1601C from New Imaging technology: VGA format 15 µm pitch) is packaged in a CLCC sample holder and electrically connected to it using bonding connections. The latter are then buried within an insulating resist which mechanically prevents bonding damage during later spin-coating step and that prevents formation of electrical short during NC deposition. The ROIC is then cleaned in a $O_2$ plasma cleaner for 10 min. The ROIC is then placed on a spin-coater and 30 µL of HgTe ink is spin-coated at 1000 rpm and an acceleration of 500 rpm.s$^{-1}$ for 60 s and a drying step at 2000 rpm (1000 rpm.s$^{-1}$) for 120 s. The film is further dried for 2 hours under primary vacuum. The ROIC is then operated on a special mode where the bias is applied to one pixel out of every two, arranged in a checkerboard configuration, enabling application of electric field along the substrate. Although this process reduces the resolution by half, it is possible to obtain a full VGA resolution by taking images in both checkerboard configuration and then combining them.

## 2. Importance of the core on the shell morphology

Initial attempt to grow CdS shell have been conducted using cores obtained from the core growth described by Keuleyan *et al.*[4] which is not known to lead to spherical shape particles[5] with a serious degree of aggregation. Then, using the procedure developed for HgSe[1] by Kamath *et al.* to grow CdS shell, we obtain systematic strong redshift of the spectrum (**Figure S 1**) and weak colloidal stability for the final material. Same situations were obtained while attempting to start with even smaller cores where final band edge was always above 2.5 µm.

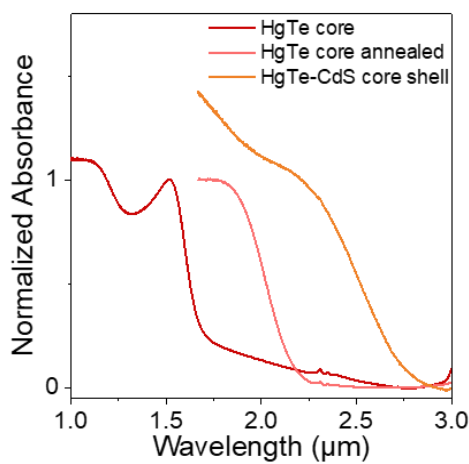

*Figure S 1 Core-shell material obtained from non-spheric core. Absorption spectra for core-only, core-annealed at the CdS growth temperature and HgTe-CdS core-shell.*

To solve this problem of colloidal stability and favor homogeneous shelling of the material we then switch the core growth procedure to a seeded growth strategy as proposed by Prado *et al.*[6], which leads to more sphere-shaped particles with reduced aggregation as shown in **Figure S 2**a. Using the Kamath *et al.* procedure for shell growth on such cores leads to a clear growth of additional material with the formation of long arms, see **Figure S 2**b. Such arms are very unfavorable to achieve densely packed film and thus poorly suited to obtain strong absorption and large mobility. XRD unveiled that the used conditions favor the growth of wurtzite CdS. This is why in the main text the procedure have been updated and includes $CdCl_2$ which can act both as Cd source but also as ligand[7,8] to favor the formation of zinc blende CdS.

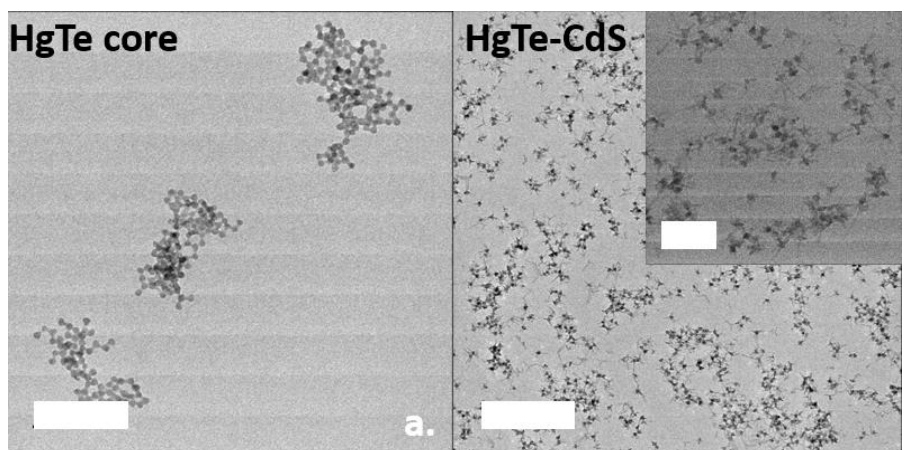

*Figure S 2 Microscopy from core-shell using aggregated non spherical cores*. a. TEM image of HgTe core obtained using Prado's procedure. [6] Scale bar is 100 nm. b. TEM image of HgTe-CdS core shell. Scale bar is 200 nm. The inset is a zoom on same material. Scale bar is 50 nm.

## 3. Characterization of the core-shell material

While growing the shell, the presence of CdS can be evidenced. In the visible absorption spectrum (**Figure S 3**a), a small feature at high energy (360 nm or 3.44 eV) corresponding to confined form of CdS is observed. Infrared spectra of core and core-shell (**Figure S 3**b) reveal a limited shift (- 25 meV) resulting from the shell growth. This enables to maintain the short-wave infrared band edge absorption of the material and prevents excessive increase of the dark current which may have resulted from band gap reduction. The Raman data (**Figure S 3**c) clearly highlights a feature relative to CdS (LO mode at 290 cm$^{-1}$) in the core-shell material spectrum. X-ray diffraction of core-shell material (**Figure S 3**d) shows in addition of the features from HgTe zinc blende, at around 2θ=26° a peak that we can attribute to the (111) direction peak from CdS. We also note a shift of the (111) peak from HgTe toward wider angle meaning that the HgTe lattice get constrained upon shell growth.

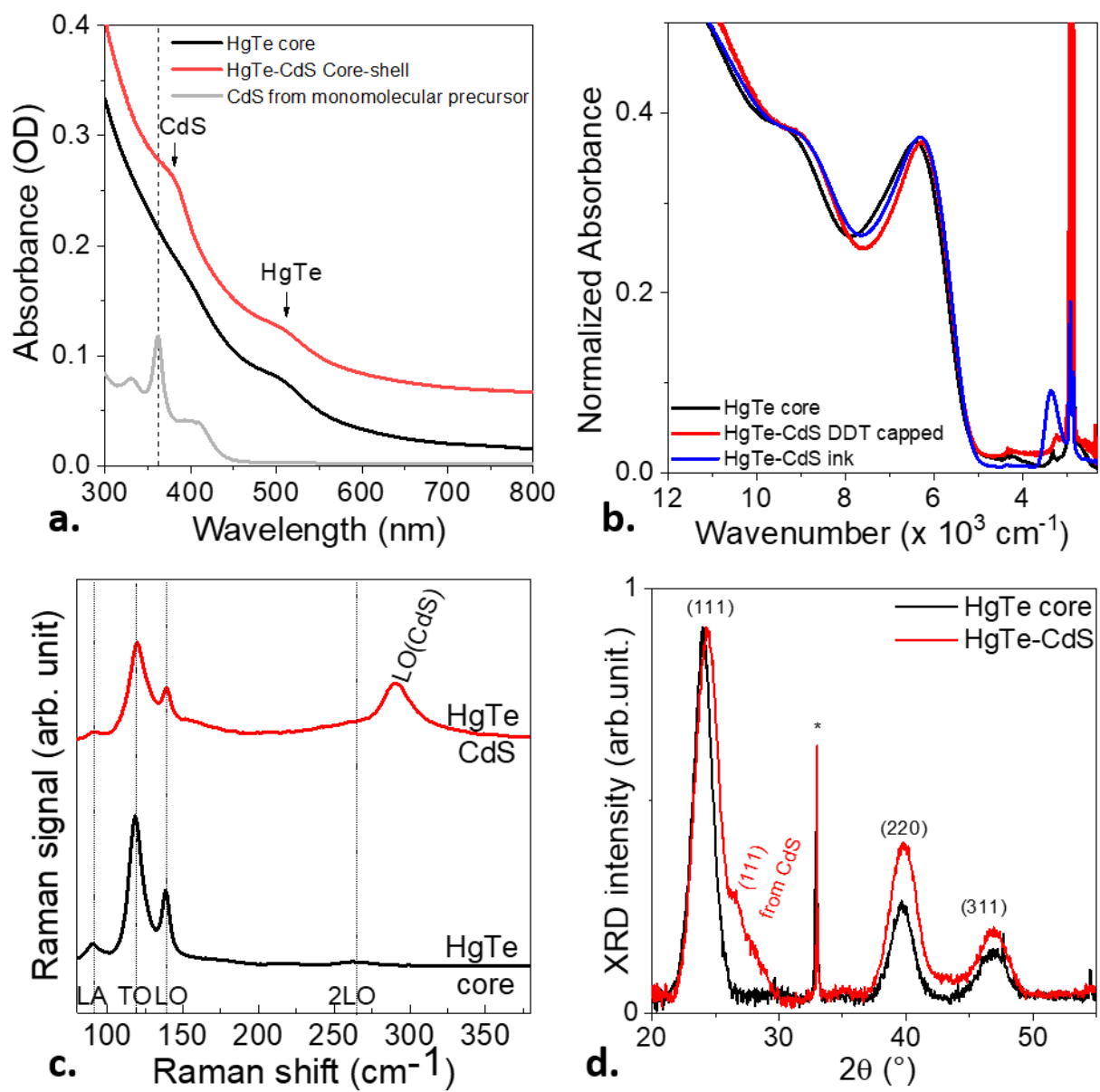

*Figure S 3 Spectroscopic and strctural characterization of HgTe core and HgTe-CdS core shell NCs.* a. UV-visible absorption spectra from HgTe core and HgTe-CdS core shell NCs as well as from the material obtained from decomposition of cadmium bis(phenyldithiocarbamate).[1] b.

*Infrared absorption spectra from HgTe core and HgTe-CdS core shell NCs with DDT capping and under ink form. c. Raman spectra of HgTe core and HgTe-CdS core shell NCs, highlighting the presence of LO mode from CdS in the core shell. d. X-ray diffractogram from HgTe core and HgTe-CdS core shell NCs.*

From EDX we determine the Hg/Te ratio to be around 1.25 indicating an Hg excess. The Hg/Cd ratio is found at around 1.7. If we assume a sphere shape for the particle with an initial diameter of 4.4 nm as revealed from TEM, we determine a shell thickness of 0.27 nm corresponding to ≈1 ML of CdS.

X-ray photoemission measurements (**Figure S 4**) further confirms the growth of CdS materials. The survey spectrum from the pristine HgTe cores (**Figure S 4a**) shows the presence of Hg and Te features in addition of C, Cl and S. The latter three are coming from the ligands ($HgCl_2$ and thiol) used to prepare the ink film. Note that we cannot notice any features coming from O 1s (binding energy of 532 eV), in spite of air preparation of the film and in absence of high temperature degassing step. This confirms the weak tendency of HgTe for oxidation.[9]

After shell growth (**Figure S 4b)** contribution of Cd (Cd 3d at 405 eV binding energy, **Figure S 4e**) and S (S2p at around 162 eV binding energy, **Figure S 4f**) are visible on the spectrum.

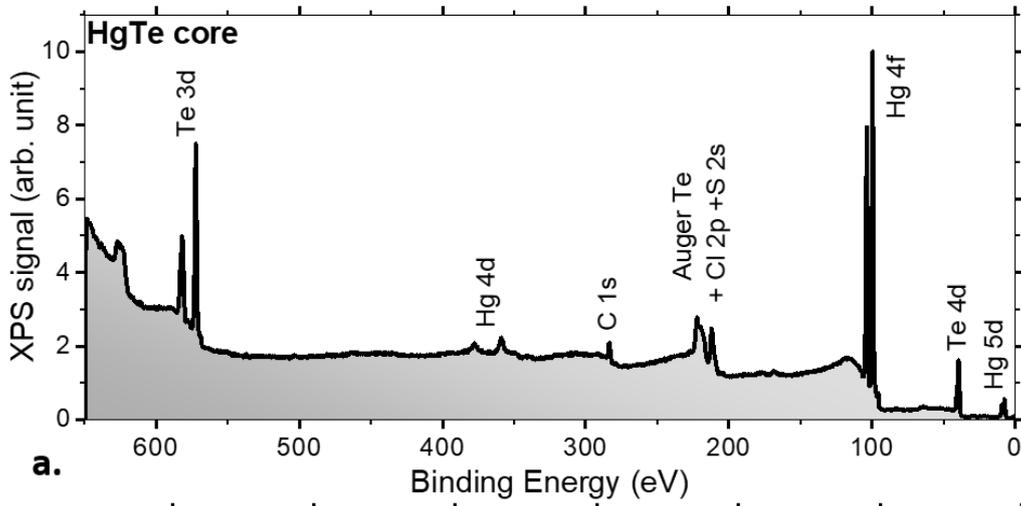
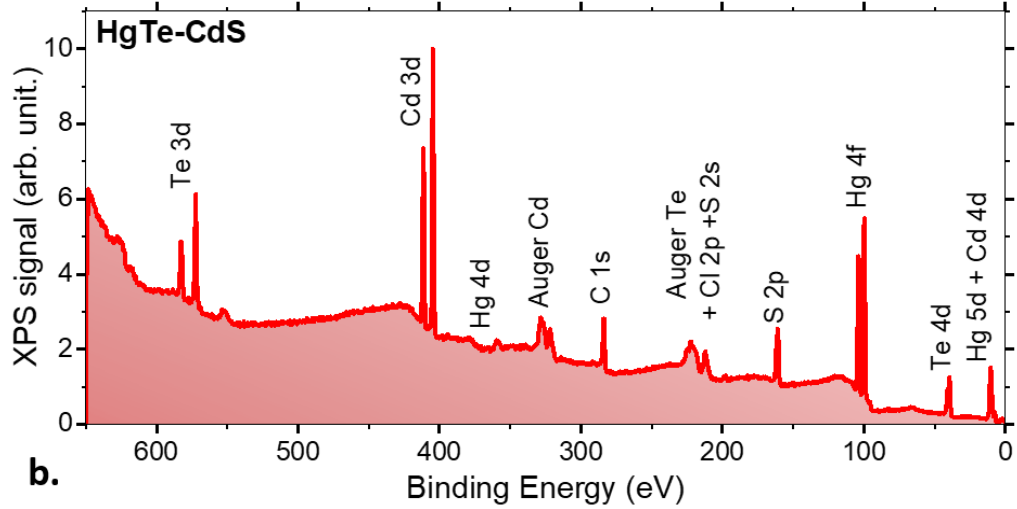
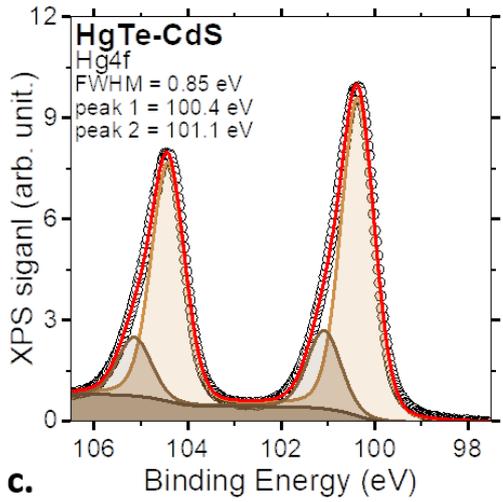
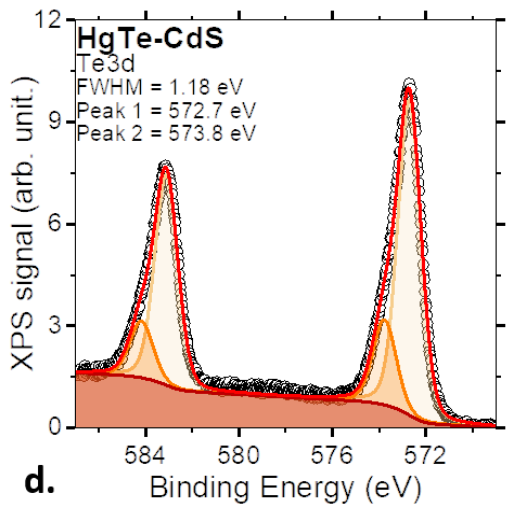
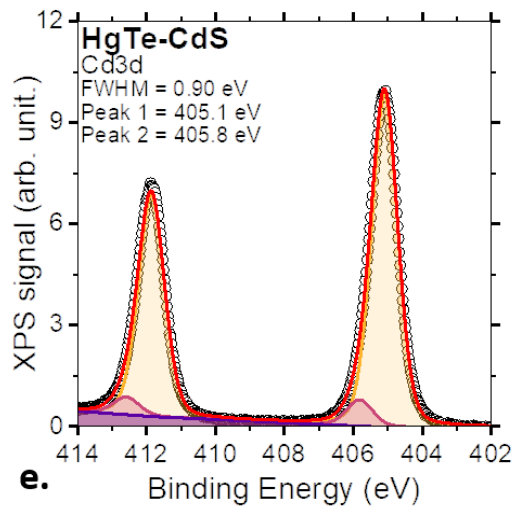
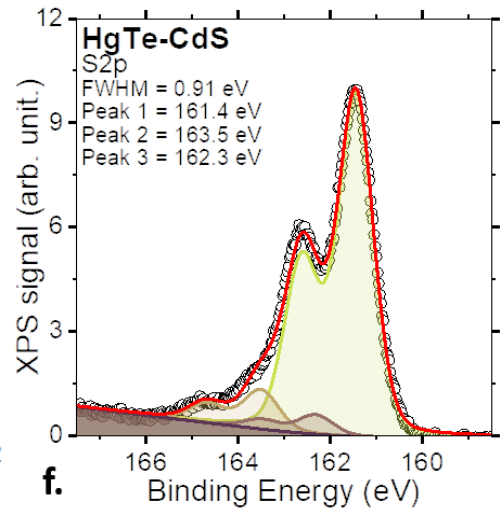

S10

*Figure S 4 XPS analysis of HgTe core and HgTe-CdS core-shell NCs*. a. Survey spectrum for HgTe cores processed as ink (thiol+$HgCl_2$ on surface) acquired for hν=700 eV. b. Survey spectrum for HgTe-CdS core-shell processed as ink (thiol+$HgCl_2$ on surface) acquired for hν=700 eV. part c to f are respectively the high resolution photoemission spectra of the Hg 4f, Te 3d, Cd 3d and S 2p states.

Recently, Alchaar *et al.* have noticed that HgTe NCs in contact with certain metals and in particular Ag, tend to form amalgams.[10] This is an obvious problematic situation for the design of diodes in which the absorbing layer is generally surrounded by metals that are possibly not noble metals and but also for the coupling with read out integrated circuit where gold is commonly excluded because of its tendency to form deep trap in silicon. A clear signature of this transformation is the observation of reduced Hg (i.e. metallic mercury as opposed to $Hg^{2+}$ in HgTe NCs[10]). With core shell HgTe-CdS, as shown in figure **Figure S 5**, the spectrum from the Hg 4f remains the same coupled to gold (no amalgam) or to silver (that forms amalgams with pristine HgTe). Another benefit of the shell thus relates to the increased chemical stability.

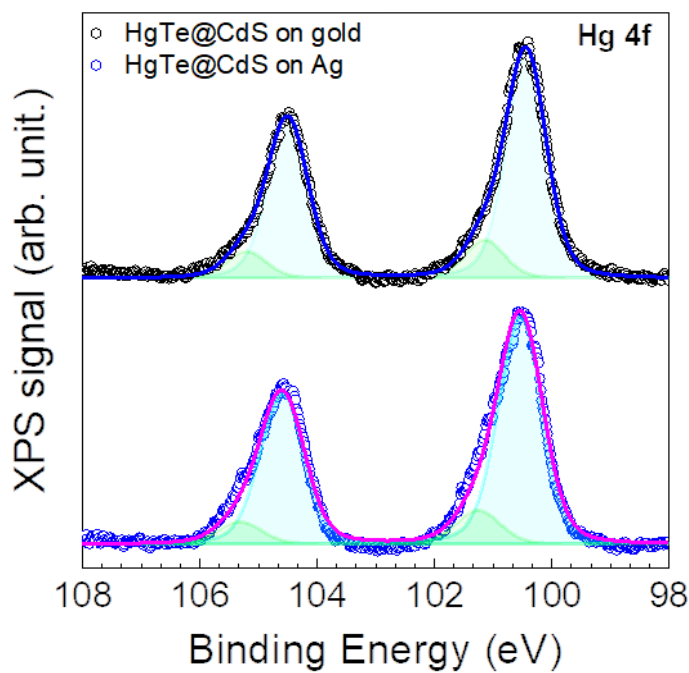

*Figure S 5 Chemical stability toward amalgam formation*. Hg 4f states photoemission spectra for HgTe-CdS core-shell deposited on gold (top) and silver (bottom) acquired with a photon energy of 700 eV. The main component in the spectra is related to the Hg-Te bond. The small contribution at higher binding energy is related to Hg atoms bound to thiols.

For sake of completeness, we have also measured the optical index of the core-shell that can be a central input for their future integration into photonic structures. Compared to HgTe with the same surface chemistry, the refractive index appears at around 2.05 that is weaker to what is observed for pristine HgTe (n=2.2-2.3 range[11]). Though the effect may be attributed to the presence of CdS, difference in NC packing may also be involved.

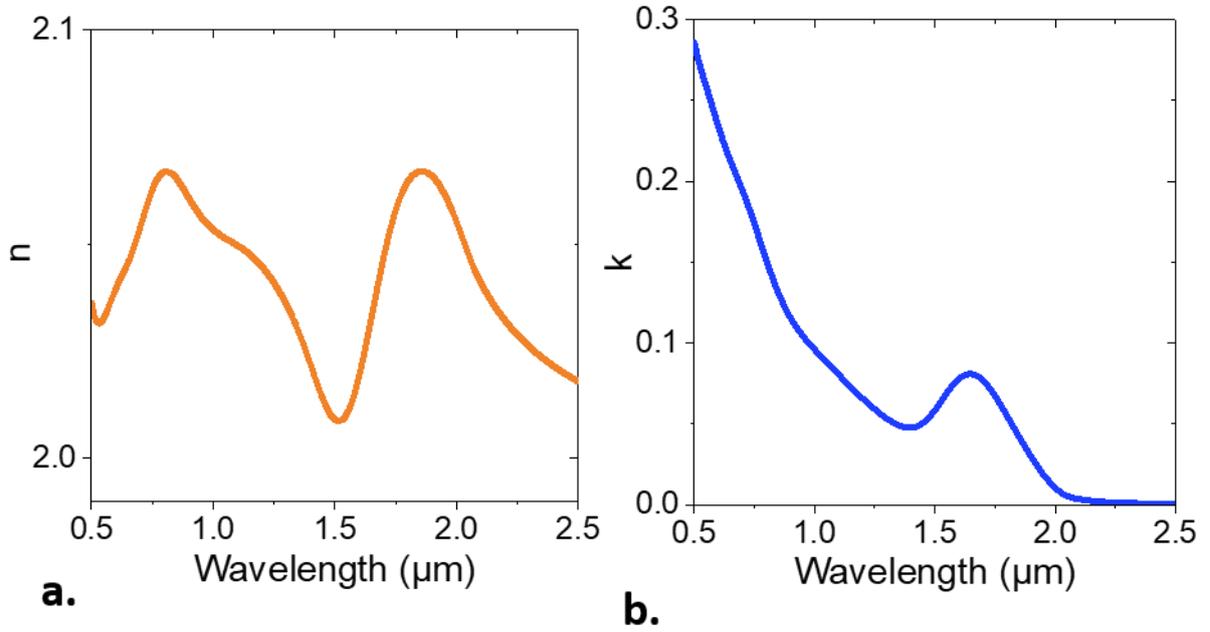

*Figure S 6 Complex optical index of thin HgTe-CdS NC film. a. Spectrally resolved refractive index. b. Spectrally resolved extinction coefficient.*

## 4. Thermal conductivity determination

**Thermal conductivity measurements**: The thermal conductivity of the nanocrystal films was obtained using modulated thermoreflectance (MTR) microscopy. In this experiment, the pump beam (532 nm, Cobolt MLD) was scanned in a line across the sample and absorbed directly by the nanocrystal film while the resulting change in reflectance was probed using a fixed probe beam (488 nm, Oxxius). Films of HgTe/CdS core/shell and HgTe core-only nanocrystals were deposited onto borosilicate substrates and measured to have thicknesses of 200 nm by a profilometer (Dektak 150 Veeco). The intensity of the pump beam was modulated by an acousto-optic modulator at a frequency $f$ and the beam was focused on the sample with an objective lens (N.A. = 0.5). The change in reflectance of the probe beam was measured using a photodiode and a lock-in amplifier to record the AC reflectivity component in a frequency range between $f$ = 50 and 500 kHz. Pump beam powers of 35 and 22 µW and probe beam powers of 120 and 40 µW were used for the core/shell and core-only samples, respectively. The experimental amplitude and phase profiles were fit according to numerical simulations based on Fourier's law to extract the thermal conductivity of the films using a 1.5 µm Gaussian pump beam diameter and diffraction-limited Airy probe profile.[12,13] A detailed procedure of the data treatment can be found from ref [14]. The thermal conductivities of the HgTe/CdS core-shell and HgTe core-only NC films were found to be 1.0 ± 0.2 and 0.9 ± 0.2 W/m·K, respectively, see **Figure S 7**.

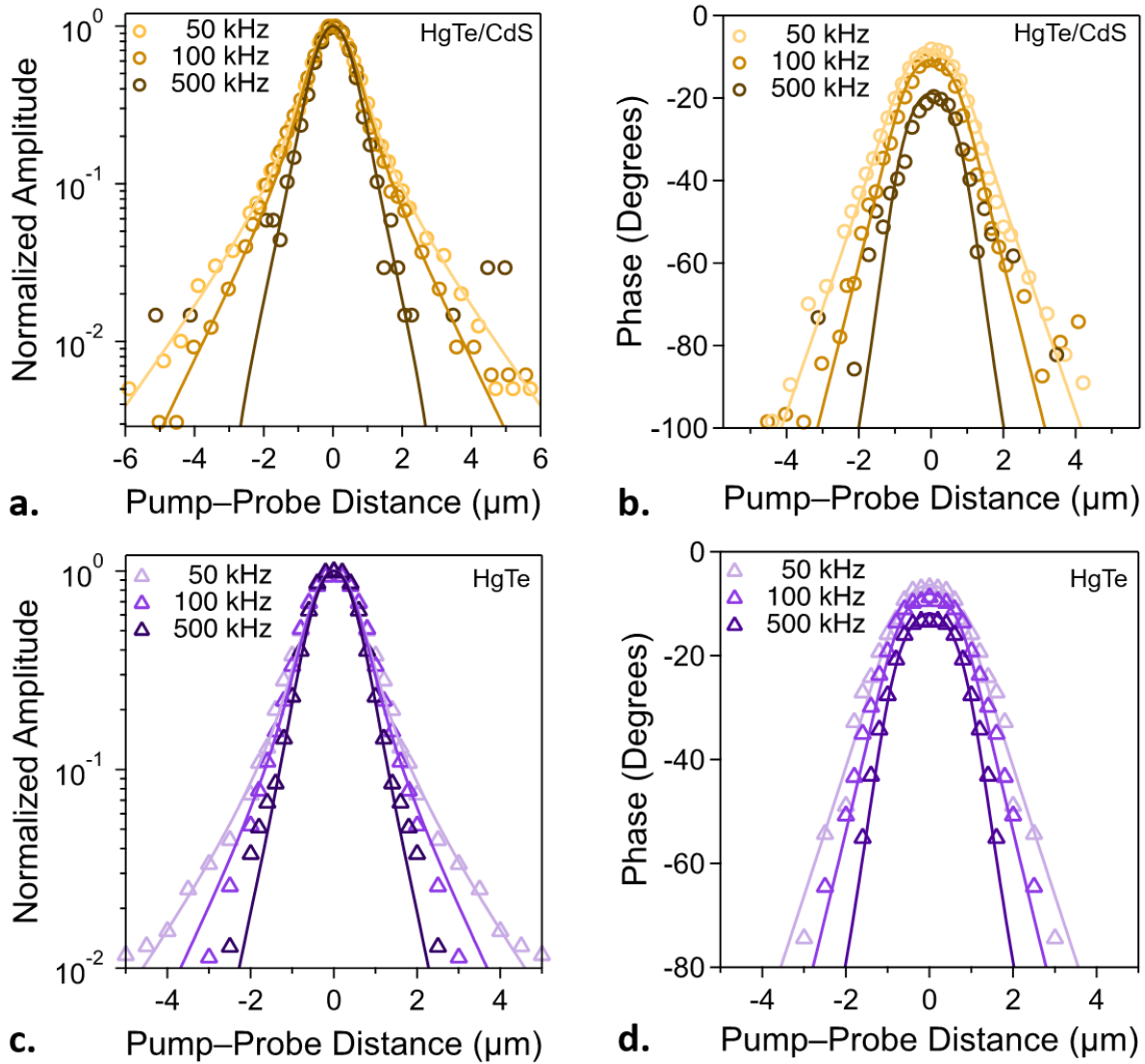

**Figure S 7 Thermal conductivity measurements**. a. Amplitude and (b.) phase of modulated thermoreflectance (MTR) signal collected for a HgTe/CdS core-shell film. c. Amplitude and (d.) phase of MTR signal collected for HgTe core-only film. Data are shown with symbols while solid lines show the corresponding numerical simulations. For the HgTe core-only data, positive and negative pump–probe distances were averaged around the center to improve signal-to-noise then mirrored for clear presentation.

## 5. 8 band 3D k.p modelling and self-consistent modelling

Electron and hole states in the absence of electrostatic interactions are calculated within the framework of 8-band k·p theory, $H = H_{kp} + V_c(x,y,z)$. Here $H_{kp}$ is the multi-band k·p Hamiltonian, expanded on a basis of $\Gamma_6$, $\Gamma_8$ (heavy and light hole subbands), and $\Gamma_7$ (split-off) bands, including spin-orbit interaction. The basis set, resulting Hamiltonian and HgTe material parameters are taken from Ref. [15], but we introduce the momentum operator (rather than wave vector) to describe quantum confinement in all three dimensions. It is worth noting that the material parameters include valence band warping (beyond the spherical approximation). We consider HgTe parameters (masses and Kane parameter) all over the heterostructure, since including CdS masses leads to the appearance of spurious states. The presence of the CdS layer and the insulating environment is accounted for through the confining potential $V_c$. The latter is defined by the band offset between conduction ($\Gamma_6$) or valence ($\Gamma_8$ and $\Gamma_7$) bands in each material. The schematic below shows the $V_c$ along a radial section of the heterostructure. At the HgTe/CdS interface, the values are inferred from the valence band offset calculated by first-principles methods (VBO=1.35 eV)[16] along with bulk band gaps ($E_g^{HgTe}$=-0.303 eV, $E_g^{CdS}$=2.5 eV).[17] For the ligands, we simply assume a large confinement potential (8 eV), as suggested by the large HOMO-LUMO gap of ligands, symmetrically distributed between conduction and valence bands. Hamiltonian $H$ is integrated by finite elements method in 3D cartesian coordinates using Comsol 6.1.

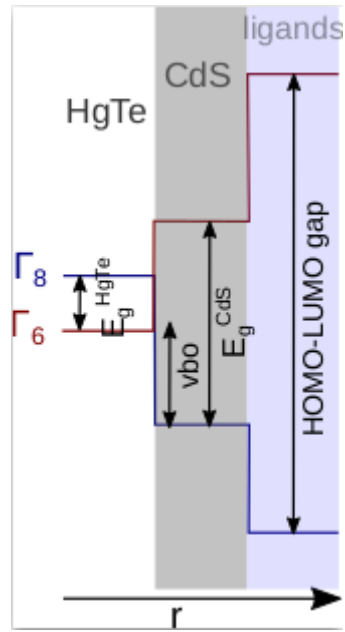

***Figure S 8 Radial band profile of the HgTe/CdS/ligand heterostructure in the simulations.***

Exciton states, with full inclusion of electrostatic interactions, is calculated by:

$$H_X = H(r_e) + H(r_h) + \Sigma(r_e) + \Sigma(r_h) + V(r_e, r_h).$$

Here, $r_e$ is the coordinate of the lowest electron state (first state above the Fermi level). Likewise, $r_h$ is the coordinate of the chosen hole state (first or second state below the Fermi level, $P_{3/2}$ and $S_{3/2}$ states, respectively). Σ is the self-energy term arising from the interaction of carriers with the surface charge they induce on the dielectrically mismatched interface, which are calculated using expressions for spherical structures.[18] Divergences are avoided by replacing the abrupt interface profile by a cosenoidal one of width δ=1 Å. A static dielectric constant $\varepsilon_{QD}$=21 is used for the quantum dot (bulk HgTe value), and $\varepsilon_{out}$=3 for the surrounding organic medium (unless otherwise noted). The latter value is suggested by recent measurements.[19] $V(r_e,r_h)$ is the Coulomb interaction between electrons and holes. It is obtained by integrating the Poisson equation in a dielectrically inhomogeneous medium, using Comsol 6.1. Hamiltonian $H_X$ is solved through a self-consistent iteration process.

Special care needs to be taken for the evaluation of Σ terms, since the self-energy potential presents sharp variations near the interface, which are difficult to capture in a 3D finite element mesh. To tackle this problem, we first solved the Hamiltonian in the absence of Σ. The resulting (3D, cartesian) charge density was then used to construct the radial density $\rho(r)$ upon integration over the angles. Next, the radial density was integrated with the spherically symmetric self-energy potential $\Sigma(r)$. A non-uniform mesh was adopted to cover for the abrupt dependence of the self-energy near the dielectric constant discontinuity. The self-energy was determined as $\langle \Sigma \rangle = \int r^2 \, \rho(r) \, \Sigma(r) \, dr = \sum_n l_n$, with $l_n = \int_{r_n}^{r_{n+1}} r^2 \, \rho(r) \, \Sigma(r) \, dr$. For the evaluation of the integral $l_n$, the product $\rho(r)\Sigma(r)$ was replaced by its linear interpolation taken at the *n-th* integral limits. The integral of the interpolating function with the Jacobian $r^2$ was calculated analytically.

## 6. Shell induced pressure

The shell growth results in strain applied on the core due to the lattice mismatch between HgTe and CdS (10 %, see **Table S 1**). A direct signature of this pressure is evidence from the X-ray diffractogram that presents a shift toward wider angles of the HgTe pattern in the core-shell structure, see **Figure S 3**d. Ithurria *et al.* developed a continuous mechanical model giving an analytical expression for the estimation of shell induced pressure.[21]

$$P_0 = \frac{2.E_{core}.E_{shell}.\varepsilon.(\frac{R_{shell}^3}{R_{core}^3} - 1)}{(2E_{shell}(1 - 2\nu_{core}) + E_{core}(1 + \nu_{shell}))\frac{R_{shell}^3}{R_{core}^3} - 2(E_{shell}(1 - 2\nu_{core}) - E_{core}(1 - 2\nu_{shell}))}$$

*(Equation 1)*

In this expression $E$ is the Young modulus, $\nu$ the Poisson coefficient, $R$ the particle radius and $\varepsilon$ the deformation *($a_{core}$-$a_{shell}$)/ $a_{core}$* with *a* the lattice parameter.

***Table S 1 Mechanical parameters from bulk HgTe and CdS*** *according to ref[20].*

|  | HgTe | CdS |
|---|---|---|
| **Young modulus (GPa)** | 34 | 68 |
| **Poisson coefficient** | 0.288 | 0.34 |
| **Lattice parameter (nm)** | 0.6453 | 0.582 |

Assuming a 4.2 nm for the HgTe core diameter and a 1ML (≈0.3 nm) shell of CdS, we can determine from **Figure S 9** a 1.7 GPa pressure. This value is higher than the expected threshold (1.4 GPa) for the zinc blende to cinnabar transition in bulk HgTe,[22,23] which may have dramatic consequences since the structural phase change also come with an electronic phase change from semimetal to narrow band gap semiconductor and thus will have deeply affected the optical feature. However, Livache *et al.* have shown that the low-pressure phase is stabilized over a broader phase for HgTe under a NC form at least up to 3 GPa.[20,24,25]

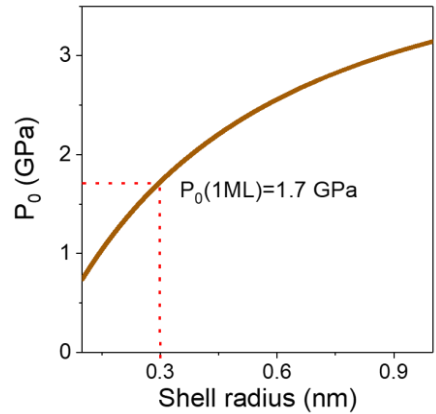

***Figure S 9 Pressure induced by the shell on the core*** *according to equation 1.*

Though the low-pressure phase may be maintained, the pressure is still expected to lead to a blue shift of the spectral response of around 60 meV/GPa, thus leading to a blue shift of around 100 meV while we grow a 1 ML thick shell that imply on the core 1.7 GPa. Experimentally, we observe a small 25 meV redshift as shell is grown, which may result from a competition between a redshift as confinement gets reduced and a pressure induced blue shift.

## 7. Carrier mobilities with and without shell

As it may have been anticipated the growth of shell, leads to improve thermal stability but is paid though a reduced carrier mobility, see **Figure S 10**, which typically drop by two orders of magnitude.

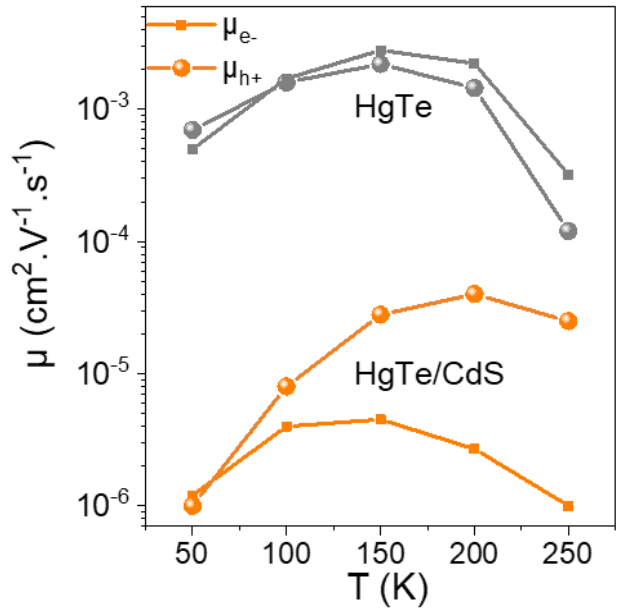

*Figure S 10 Electron and hole mobility for core and core shell NCs film.*

## 8. Performances and aging of the camera based on core-shell NCs

To determine the external quantum efficiency (EQE) of the HgTe-CdS core-shell camera we proceed by comparison with an InGaAs sensor which EQE is known (90%, see **Figure S 11**). For both systems we integrate the amplitude of the signal (in adu: analog digital unit) as a function of power for a given integration time. Then, the comparison of the slope is equal to the ratio of the EQE.

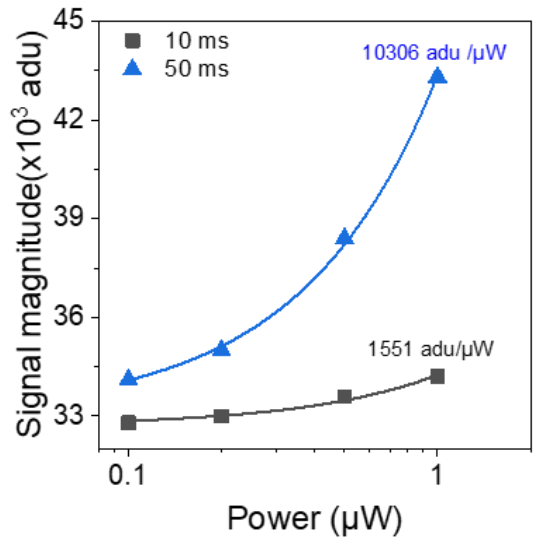

*Figure S 11 Procedure to determine EQE. Signal for the InGaAs camera presenting 90 % EQE as a function of the incident power at 1.55 µm for two integration times.*

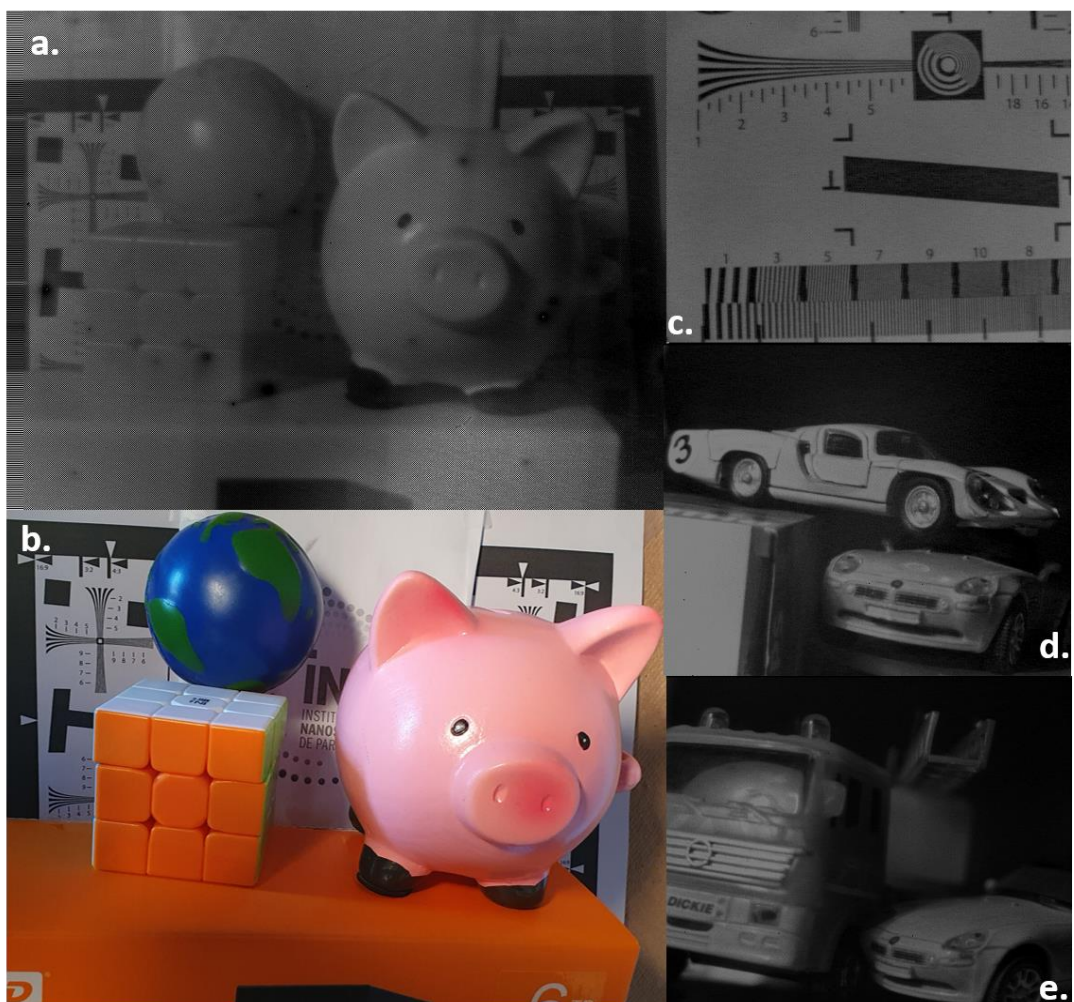

*Figure S 12 Additional infrared images acquired with the imager based on HgTe-CdS NCs as absorbing layer. a. (resp b.) SWIR (resp visible) images. c-e are other examples of SWIR image acquired with the imager based on HgTe-CdS NCs as absorbing layer.*

**Figure S 13** illustrates the processing of the image that are corrected with a two-point correction. A dark image is used to remove dark current and an image under homogeneous illumination is bused to generate a gain map.

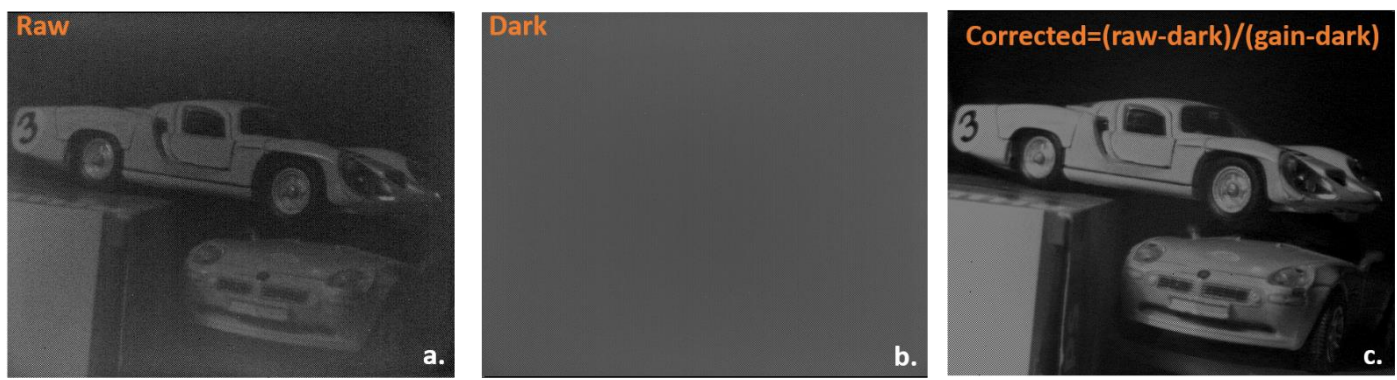

***Figure S 13 Image correction processed***. *For a given acquisition time we systematically acquire a dark uniform image (called dark) as well as an image under a homogeneous illumination (called gain). Then the corrected images are obtained according to the following treatment corrected=(raw-dark)/(gain-dark)*

**Figure S 14** compares the aging of the imager for which the active layer is made of HgTe core and HgTe-CdS core-shell NCs. With core-only, dark current rises quickly which prevent long integration time. the latter can typically drop from ms (without cooling) to few tens of µs the next days. A situation that contrasts with the core-shell for which ms integration time can be maintained after 4 months in air and 2 days of continuous operation.

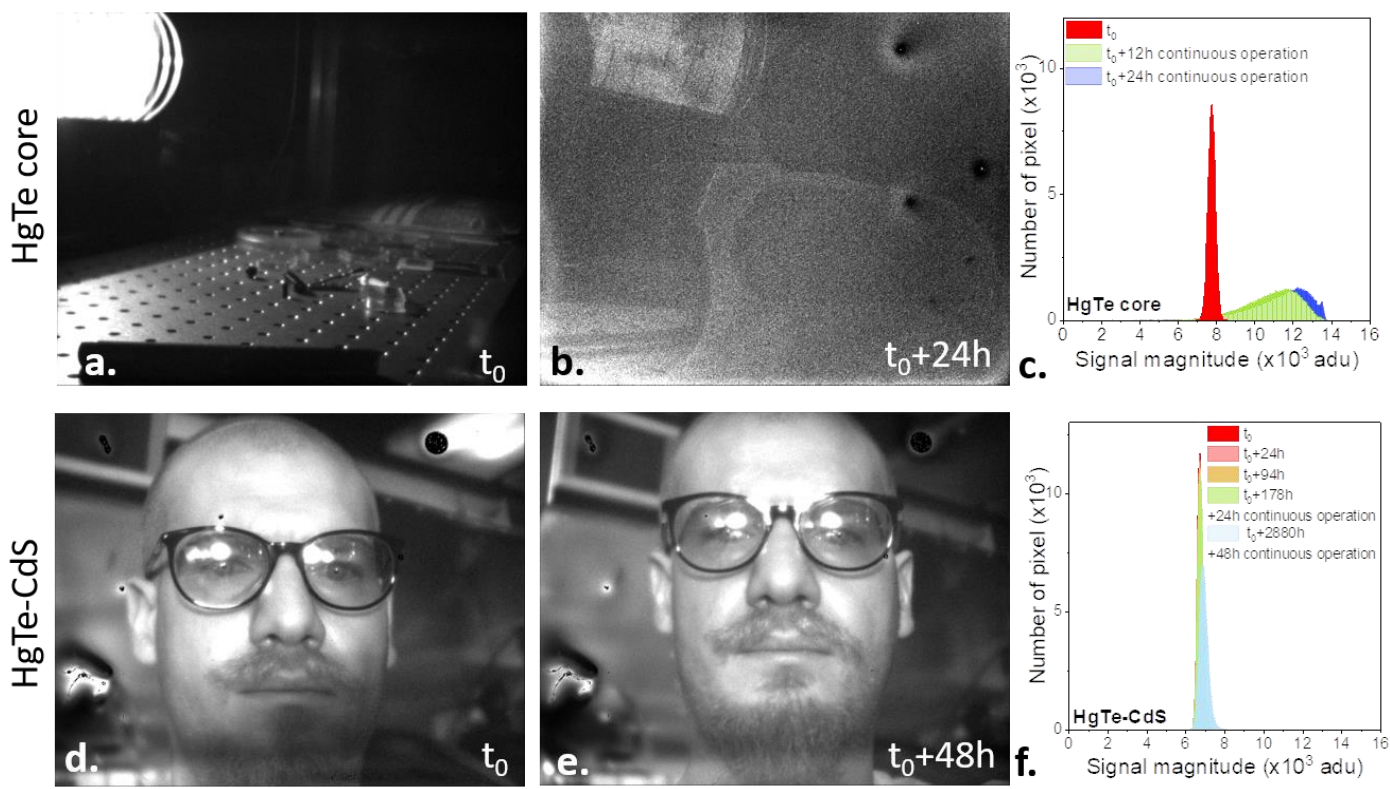

***Figure S 14 Aging of HgTe and HgTe-CdS imager***. *a. (resp b.) Image from HgTe NC based imager after fabrication (resp after 24 h continuous operation). c. Dark current histograms at different times for the HgTe NC based imager. d. (resp e.) Image from HgTe-CdS NC based imager after fabrication (resp after 48 h continuous operation). f. Dark current histograms at different times for the HgTe-CdS NC based imager. Device is operated at room T without cooling system.*

We already observe improved chemical stability of the core-shell by the lack of amalgam formation. We here test the potential to sustain a step of atomic layer deposition (a common procedure to package semiconductors. Previous attempt on HgTe NCs have led to a complete disappearing of the photoresponse.[26] Here, conductive and photoconductive properties remain mostly unchanged after 5 h of ALD leading to the deposition of 5 nm of $Al_2O_3$.

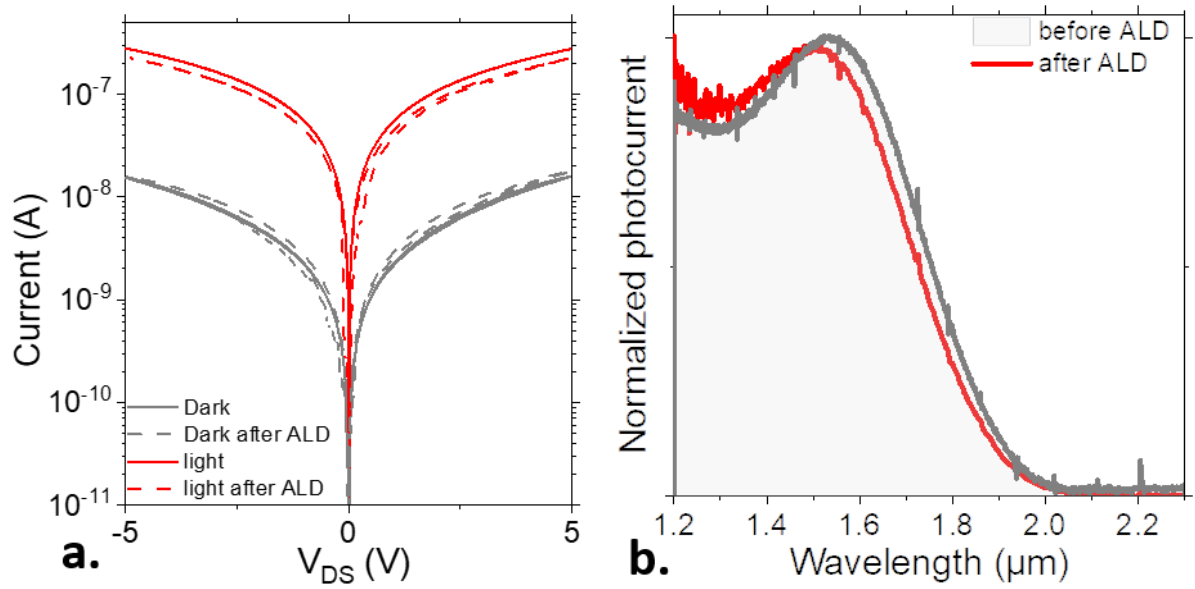

*Figure S 15 Performance after ALD*. a. IV curves of HgTe-CdS NC based film in the dark and under illumination (4 mW at 1.55 µm) for the device before and after ALD deposition (5 nm of $Al_2O_3$ deposited at 50 °C over 5h). b. Photocurrent spectra for the HgTe-CdS NC based film before and after ALD.

# 9. Supplementary references